\documentclass[prb,reprint,letterpaper,eprint,longbibliography,footinbib,notitlepage,floatfix]{revtex4-1} 

\usepackage{pdfpages} 
\usepackage{pgffor} 

\makeatletter
\AtBeginDocument{\let\LS@rot\@undefined}
\makeatother

\def\supplementfilename{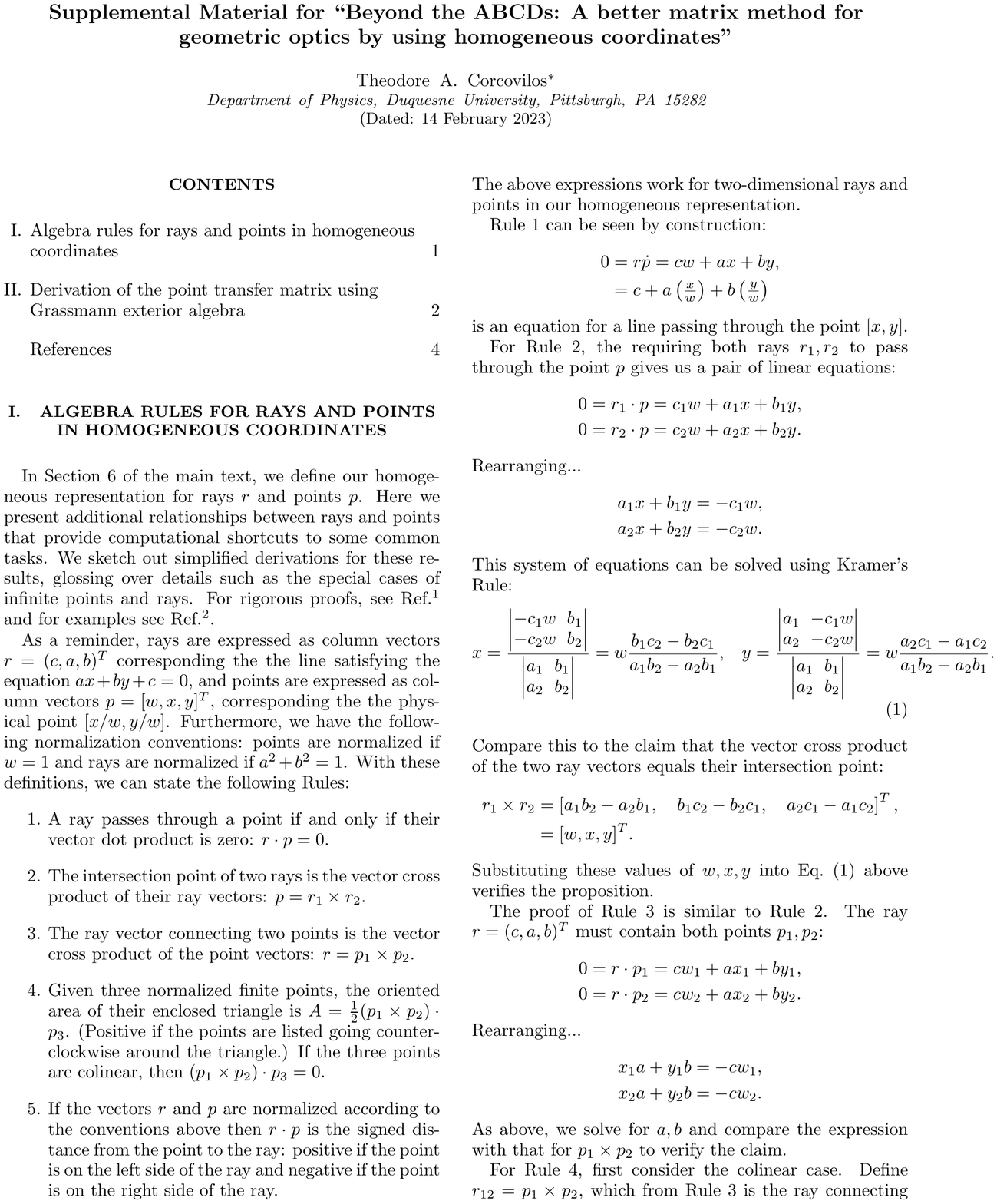}

\pdfximage{\supplementfilename}
\def\numbersupplementpages{\the\pdflastximagepages}

\newif\ifarXiv
\arXivtrue 

\usepackage[colorlinks, allcolors=blue]{hyperref}

\usepackage{amsmath}  
\usepackage{amsfonts} 
\usepackage{graphicx} 

\begin{document}
\title{Beyond the ABCDs: A better matrix method for geometric optics by using homogeneous coordinates}
\author{Theodore A. Corcovilos}
\email{corcovilost@duq.edu}
\affiliation{Department of Physics, Duquesne University, Pittsburgh, PA 15282}
\date{14 February 2023}
\begin{abstract}
Geometric optics is often described as tracing the paths of non-diffracting rays through an optical system.
In the paraxial limit, ray traces can be calculated using ray transfer matrices (colloquially, ABCD matrices), which are $2\times2$ matrices acting on the height and slope of the rays.
A known limitation of ray transfer matrices is that they only work for optical elements that are centered and normal to the optical axis.
In this article, we provide an improved $3\times3$ matrix method for calculating paraxial ray traces of optical systems that is applicable to how these systems are actually arranged on the optical table: lenses and mirrors in any orientation or position (e.g.~in laboratory coordinates), with the optical path zig-zagging along the table.
Using projective duality, we also show how to directly image points through an optical system using a point transfer matrix calculated from the system's ray transfer matrix.
We demonstrate the usefulness of these methods with several examples and discuss future directions to expand applications of this technique.
\end{abstract}
\maketitle

\section{Introduction}
Geometric optics describes light as non-diffracting rays traveling through media and surfaces using the laws of reflection and refraction.
Assuming the optical system of lenses, mirrors, etc., has a rotational axis of symmetry (the optical axis) we can define the paraxial approximation as the limit where the ray height $h$ relative to the axis is much smaller in magnitude than other lengths in the system, such as radii of curvature or focal lengths, and the slope $m$ the ray makes with the optical axis has a magnitude much smaller than one.
In this paraxial limit, the effect of optical elements such as lenses and mirrors may be approximated by linear functions of a ray's height and slope.
This approximation is often the first step in designing optical systems and is also one of the first optics topics presented to students (yielding the Gaussian and Newtonian image equations, the lensmaker's equation, etc.).

Under these conditions, the laws of reflection and refraction at planar and spherical surfaces may be replaced by their Taylor series approximations to first order in $h$ and $m$, and therefore may be expressed as matrix equations.\cite{halbach1964,gerrard1994,Pedrotti-chapter,Hecht-chapter}
These matrices are known as ray transfer matrices (RTM), or ``ABCD matrices'' after their typical parameterization.
In this article, we revisit the ABCD matrices from a geometric perspective, rather than a strictly algebraic one, and use geometric insight to expand the functionality of our matrix representation.
We do this by introducing a homogeneous coordinate representation of lines, showing that the ray transfer matrices are already applicable in this system, and then adding geometric transformations to our set of allowed operations.
The ideas we present here are implicit in advanced treatments of optics, for example, the geometric optics chapters of Born and Wolf,\cite{Born1999} but have not been previously fleshed out into an easy-to-use form.
We fill in the missing steps, and
the end result is a set of relatively simple algebraic rules for modelling almost any common optical setup.

The key new calculational tool we use, expressing lines and points in homogeneous coordinates (defined in Sec.~\ref{sec:rtm_homo}), is well known in the computer graphics community where it is used to express rotations, translations, affine transformations (e.g.~shears), and perspective transformations as matrices.\cite{pharr2016}
Homogeneous coordinates are also a natural setting for projective geometry (particularly \emph{oriented} projective geometry\cite{stolfi2014}) which was a common mathematical description of geometric optics from the Italian Renaissance\cite{coxeter2003} up until the mid-1900s.\cite{cambi1959} Unfortunately, work in this area was largely abandoned when digital computers became viable for full, non-paraxial ray tracing using the exact laws of reflection and refraction.\cite{feder1963,wynne1963}

In our work, we express rays as oriented lines in homogeneous coordinates and identify the matrices that correspond to common optical elements favoring direct ray tracing calculations using the laws of reflection and refraction.
This takes us beyond the augmented ABCD matrices used by some authors\cite{gerrard1994,arnaud1976,shaomin1985,siegman1986,tovar1995,lin2006,lin2009,Lin2014} to a fully consistent geometric treatment.
A better understanding of the geometry underlying the ray transfer matrices is also helpful in solving the inverse problem of finding a set of optical elements needed to produce a given optical transformation.\cite{liu2008,tovar1997}
Most optical modelling involves the analysis of rays through the optical system, but by invoking one other new math tool, projective duality, we expand this idea to analyzing \emph{points} as well.
In other words, we can calculate images directly without tracing rays.
We call this transformation of points the \emph{point transfer matrix}, which we obtain from an algebraic manipulation of the ray transfer matrix.

The outline of this article follows.
We begin in Section~\ref{sec:rtm} with a review of the paraxial ray transfer matrices as typically used in two-dimensional or axially symmetric optical systems.
Next we show how RTMs are a disguised form of general linear operators on the vector space of lines represented in homogeneous coordinates (Section~\ref{sec:rtm_homo}).
Because we are using homogeneous coordinates, we can also consider optical elements decentered or rotated with respect to the optical axis, which we discuss in Section~\ref{ap:B} along with an example.
Careful treatment of algebraic signs allows us to preserve the propagation direction of our rays, which we explore in Section~\ref{sec:refl} to derive orientation-preserving RTMs for reflective elements.
By requiring a coincident point and line to remain so after imaging, we find a key new result that we call point transfer matrices (PTMs) (Section~\ref{sec:ptm}), and provide some examples demonstrating how PTMs simplify imaging calculations and the analysis of optical systems (Section~\ref{sec:ex}).
Finally, we close in Section~\ref{sec:3d} with a discussion of what would be required to extend these results to three-dimensional systems and the complications that arise therein.

The Supplemental Materials\cite{SMNote} contain an alternative derivation of the PTMs using Grassmann exterior algebra (an algebra that uses different ``grades'' to represent the geometric ideas of lines, areas, and volumes), leading to a compact proof of the Scheimpflug principle of tilt-shift photography, as well as example python code for implementing our work.  In particular, the python code contains calculations for the examples in this main text and additional examples that we omitted to save space.

\section{Ray transfer matrices}\label{sec:rtm}
Before introducing our new concepts, we will briefly review the ray transfer matrices as conventionally used.
We refer readers to Refs.~\onlinecite{halbach1964,gerrard1994,Pedrotti-chapter,Hecht-chapter} for detailed derivations.
Several conventions exist in the literature regarding the ordering of elements and the implementation of the index of refraction.
We will follow the convention of the textbook by the Pedrottis.\cite{Pedrotti-chapter}

\begin{figure}
    \begin{center}
        \includegraphics[width=3.25in]{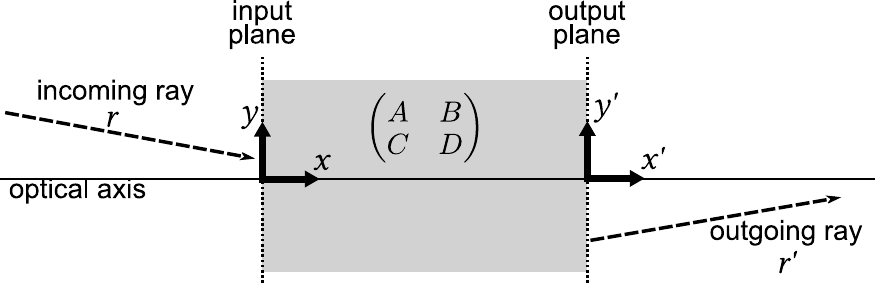}
        \caption{Ray transfer matrix.  The incoming ray ${r}$ enters the optical system (gray box) and is transformed into the outgoing ray ${r}'$ according to the ABCD matrix in Eq.~\eqref{eq:rtm}.
        The slopes of the rays are exaggerated for clarity.
        Note that the incoming coordinate system $(x,y)$ has its origin at the input plane of the system and that the outgoing coordinate system $(x',y')$ has its origin at the output of the system.
        The horizontal line represents the optical axis of the system.}\label{fig:ABCD}
    \end{center}
\end{figure}

We define an incoming ray vector ${r} = (h, m)^T$ representing a ray with slope $m$ relative to the optical axis and crossing the input plane of the optical system at height $h$.
(We use ${}^T$ to indicate matrix transposition.)
The coordinate origins of the object/image spaces are centered on the first/last surfaces of the optical system, as shown in Fig~\ref{fig:ABCD}.
The incoming ray transforms into the outgoing ray ${r}' = (h',m')^T$ through the relation
\begin{equation}\label{eq:rtm}
    \begin{pmatrix}
h' \\ m'
    \end{pmatrix} = 
    \begin{pmatrix}
        A & B \\ C & D
    \end{pmatrix}
    \begin{pmatrix}
        h \\ m
    \end{pmatrix},
\end{equation}
where $A, B, C, D$ are real-valued constants obeying $AD-BC\neq0$ such that this ray transfer matrix is invertible.
In particular, the matrix determinant is equal to the ratio of the incoming and outgoing indices of refraction, $AD-BC=n/n'$, which is often unity for common optical systems.\cite{Pedrotti-chapter}
This makes the determinant a useful check on computational results. 
The relationship described by Eq.~\eqref{eq:rtm} is shown schematically in Fig.~\ref{fig:ABCD}.
Note that some references, such as the popular textbook by Hecht,\cite{Hecht-chapter} use different conventions for the ray vectors and matrices.
A common choice is to multiply the slope in the ray vector by the index of refraction, yielding the ``reduced slope.''
We prefer to use the geometric slope to avoid complications in our transformation operators (Sec.~\ref{sec:refl}).

The ray transfer matrices are particularly useful because whole optical systems may be summarized by the product of the matrices of their components, with the sequence written from right to left (first element right-most).
If we exchange the input and output ends of the optical system (or equivalently, reverse time), the resulting ray transfer matrix is the matrix inverse of the original system RTM.

\section{Rays and ray transfer matrices in homogeneous coordinates}\label{sec:rtm_homo}
In this section we introduce a homogeneous representation of lines in a plane, which consists of a set of three coefficients, as opposed to the more familiar height and slope definition above.
This will require us to replace the $2\times 2$ ABCD matrices with $3\times 3$ ray transfer matrices.
The additional row and column provide the additional degrees of freedom that we need to implement the key results that follow in later sections.

\begin{figure}
    \begin{center}
      \includegraphics{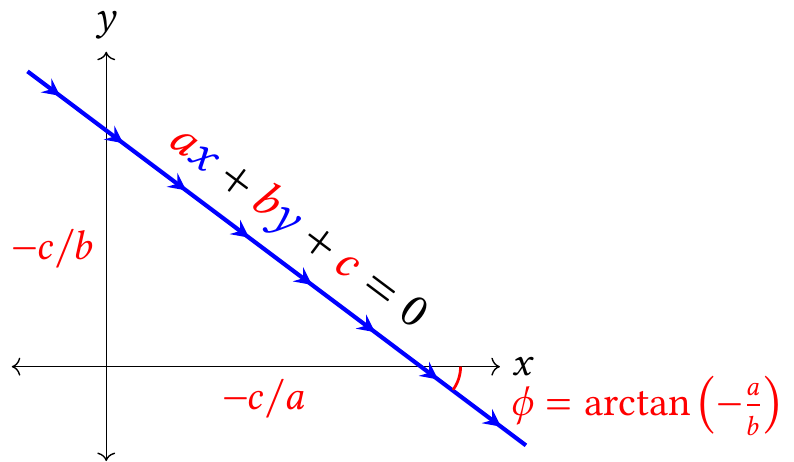}
      \caption{%
      (Color online)
        Representation of the ray $(c,a,b)^T$ as the oriented line $ax+by+c=0$.
        The coefficients of the equation can be used to calculate the $x$- and $y$-intercepts, slope, and distance from the origin.
        Note that for this particular oriented line, $a,b>0$ and $c<0$.
        \label{fig:line}}
    \end{center}
\end{figure}

Given a ray vector ${r}=(h,m)^T$, we could equivalently say that the ray falls along the line given by the equation $y=mx+h$.
It will be helpful to us to rewrite the equation for the line in the form $ax+by+c=0$ and use the coefficients $a$, $b$, and $c$ to define a vector representation of the line: 
${r}=(c,a,b)^T$. (Note the order of the coefficients, which we've chosen to facilitate comparisons of our matrices with the conventional version).
For example, our standard ray above is described by the rearranged equation $-mx+y-h=0$, corresponding to the ray vector
${r}=(-h,-m,1)^T$.
Multiplying the line equation or the vector ${r}$ by a positive scalar does not change the geometric line it represents, so this is a \emph{homogeneous} representation of the line (Fig.~\ref{fig:line}).
Although it is not necessary for our calculations, if we wish to normalize the ray vector ${r}$, a convenient choice is to require $a^2 + b^2 =1$.
With this normalization it turns out that $a$ and $b$ equal the direction cosines of the line with respect to the $x$ and $y$ axes, respectively, and $|c|$ is the perpendicular distance from the line to the coordinate origin.
If care is taken with signs during calculations (see Sec.~\ref{sec:refl} for how reflections must be handled), the rays expressed this way are oriented (have a well-defined forward direction).
For a ray $(c, a, b)^T$, the direction of propagation, measured counter-clockwise from the positive $x$-axis to the forward-going side of the ray, is the angle
\(
\phi = \arctan(-a/b), 
\)
where it is necessary to add $\pi$ radians to the value of $\phi$ in the case $b<0$ to match the forward direction of the ray.\cite{note:atan}
The propagation direction of paraxial rays in our convention is given by the sign of $b$: left-to-right rays have $b>0$ and right-to-left rays have $b<0$ (see Sec.~\ref{sec:refl} for more detail).

If the conventional $2\times 2$ ray transfer matrix of an optical system is known, the corresponding ray transfer matrix in homogeneous coordinates for nonreflecting systems (see Sec.~\ref{sec:refl} for the reflecting case) is
\[
M = {\small\begin{pmatrix}
    A & B & 0 \\ C & D & 0 \\ 0 & 0 & 1
\end{pmatrix}}.
\]
Using the height-slope definition of the ray vectors, the ray transfer equation reads
\[ {\small
\begin{pmatrix} -h' \\ -m' \\ 1 \end{pmatrix}}
=
{\small
\begin{pmatrix}
    A & B & 0 \\ C & D & 0 \\ 0 & 0 & 1
\end{pmatrix}
\begin{pmatrix} -h \\ -m \\ 1 \end{pmatrix}}.
\]
Compare this with Eq.~\ref{eq:rtm}.
The RTMs for common optical elements are summarized in Table~\ref{tab:summary}.
The additional row and column will have no apparent purpose at first, but they provide additional degrees of freedom that we will exploit later for coordinate transformations to represent decentered and rotated optical elements (Sec.~\ref{ap:B}), consistent handling of orientation (Sec.~\ref{sec:refl}), and imaging of points (Sec.~\ref{sec:ptm}).

\begin{table*}
    \caption{Summary of ray transfer and point transfer matrices\label{tab:summary}}
    \begin{center}
    \begin{tabular}[b]{p{3.0in}cc}
    Element & RTM & PTM \\
    \hline
    General form for centered elements
        & \( \begin{pmatrix} A & B & 0 \\ C & D & 0 \\ 0 & 0 & 1\end{pmatrix}\)
        & \( \begin{bmatrix} D & -C & 0 \\ -B & A & 0 \\ 0 & 0 & AD-BC\end{bmatrix}\) \\[3ex]
    Thin lens of focal length $f$
        & \( \begin{pmatrix}1 & 0 & 0 \\ -1/f & 1 &  0 \\ 0 & 0 & 1 \end{pmatrix} \)
        & \( \begin{bmatrix}1 & 1/f & 0 \\ 0 & 1 &  0 \\ 0 & 0 & 1 \end{bmatrix} \) \\[3ex]
    Free-space propagation over distance $d$
        & \( \begin{pmatrix} 1 & d & 0 \\ 0 & 1 & 0 \\ 0 & 0 & 1 \end{pmatrix} \) 
        & \( \begin{bmatrix} 1 & 0 & 0 \\ -d & 1 & 0 \\ 0 & 0 & 1 \end{bmatrix} \) \\[3ex]
    Refraction at a flat normal surface from index of refraction $n$ to $n'$
        & \( \begin{pmatrix} 1 & 0 & 0 \\ 0 & n/n' & 0 \\ 0 & 0 & 1 \end{pmatrix} \) 
        & \( \begin{bmatrix} n/n' & 0 & 0 \\ 0 & 1 & 0 \\ 0 & 0 & n/n' \end{bmatrix} \)\\[3ex]
    Refraction at a curved surface centered on the axis from index of refraction $n$ to $n'$ with radius of curvature $R$ ($R>0$ is convex)
        & \( \begin{pmatrix} 1 & 0 & 0 \\ \frac{n-n'}{Rn'} & \frac{n}{n'} & 0 \\ 0 & 0 & 1 \end{pmatrix}\)
        & \( \begin{bmatrix} \frac{n}{n'} & -\frac{n-n'}{Rn'} & 0 \\ 0 & 1 & 0 \\ 0 & 0 & \frac{n}{n'} \end{bmatrix}\) \\[3ex]
    Reflection at a plane surface
        & \( \begin{pmatrix} -1 & 0 & 0 \\ 0 & 1 & 0 \\ 0 & 0 & -1\end{pmatrix} \) 
        & \( \begin{bmatrix} -1 & 0 & 0 \\ 0 & 1 & 0 \\ 0 & 0 & -1\end{bmatrix} \)\\[3ex]
    Reflection at a spherical surface with radius of curvature $R$ ($R>0$ is convex)
        & \( \begin{pmatrix} -1 & 0 & 0 \\ 2/R & 1 & 0 \\ 0 & 0 & -1\end{pmatrix} \) 
        & \( \begin{bmatrix} -1 & 2/R & 0 \\ 0 & 1 & 0 \\ 0 & 0 & -1\end{bmatrix} \)\\[3ex]
    Translation by displacement $(u,v)$&
        \( \begin{pmatrix}1& -u & -v \\ 0 & 1 & 0 \\ 0 & 0 & 1 \end{pmatrix}\) &
        \( \begin{bmatrix}1& 0 & 0 \\ u & 1 & 0 \\ v & 0 & 1 \end{bmatrix}\)\\[3ex]
    Rotation by angle $\theta$ &
    \(\begin{pmatrix} 1 & 0 & 0 \\
         0 & \cos\theta & -\sin\theta \\
        0 & \sin\theta & \cos\theta \end{pmatrix}\) &
        \(\begin{bmatrix} 1 & 0 & 0 \\
            0 & \cos\theta & -\sin\theta \\
           0 & \sin\theta & \cos\theta \end{bmatrix}\)
    \end{tabular}
    \end{center}
    \end{table*}

\section{Coordinate transformations}\label{ap:B}
The $2\times2$ ray transfer matrices assume that the optical system has axial symmetry.
In our expanded $3\times 3$ formulation we can incorporate off-axis and rotated elements.\cite{gerrard1994,siegman1986,tovar1995}
We add to our repertoire of ray transfer matrices a matrix $R_\theta$ that rotates a ray by angle $\theta$ counter-clockwise about the origin and a matrix $T_{u,v}$ that translates a ray by $u$ in the $x$-direction and $v$ in the $y$-direction.
\begin{align}\label{eq:RTR}
R_\theta &= {\small\begin{pmatrix}1 & 0 & 0 \\
0 & \cos\theta & -\sin\theta \\
0 & \sin\theta & \cos\theta
\end{pmatrix}}, &
T_{u,v} &= {\small\begin{pmatrix} 
1 & -u & -v \\
0 & 1 & 0 \\
0 & 0 & 1
\end{pmatrix}}.  
\end{align}
Note that a translation matrix with only horizontal displacement is equivalent to the propagation matrix from Table~\ref{tab:summary}.
These coordinate transformation matrices are exact in the sense that they are not linear approximations with respect to their arguments, in contrast to the refractive sufrace matrices and spherical mirror matrix in Table~\ref{tab:summary}.

\begin{figure}
    \begin{center}
        \includegraphics{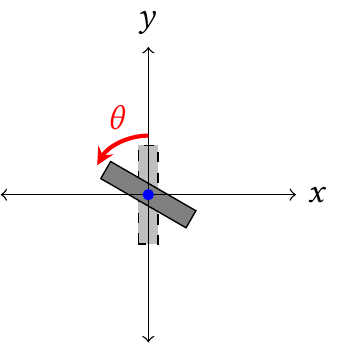}\hspace*{1cm}%
        \includegraphics{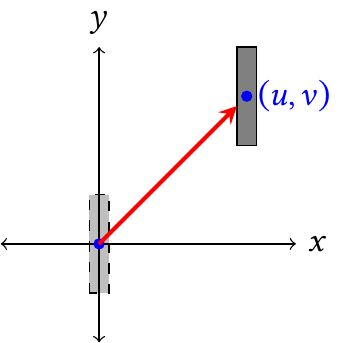}
        \caption{\label{fig:RT}%
        (Color online)
        The effects of applying the rotation operator $R_{\theta}$ (left) and translation operator $T_{u,v}$ (right) to an optical element using Eq.~\eqref{eq:transform}.}
    \end{center}
\end{figure}

The ray transfer matrices given earlier in Table~\ref{tab:summary} assume the elements are located at the origin and normal to the optical axis.
We can use the $T$ and $R$ matrices above to walk between optical elements without the restriction that they lie on the optical axis or that they are oriented normal to the original optical axis.
We can represent an element $M$ that is rotated and then translated into a new position by 
\begin{equation}\label{eq:transform1}
    M' = MR_\theta^{-1} T_{u,v}^{-1},
\end{equation}
where the outgoing coordinate axes are centered and aligned with the axis of the optical element, rather than the original optical axis.
Note that the inverses are simply $R_\theta^{-1} = R_{-\theta}$ and $T_{u,v}^{-1} = T_{-u,-v}$.
Eq.~\eqref{eq:transform1} respects the traditional application of the ray transfer matrices, where the input rays and output rays are specified in different coordinate systems.
The input rays are measured relative to a coordinate origin located at the first surface of the optical system, and the output rays are measured relative to a coordinate origin located at the final surface, as shown earlier in Fig.~\ref{fig:ABCD}.
As a simple example of Eq.~\eqref{eq:transform1}, take $M$ to be the identity matrix and then translate along the axis by a distance $d$.  The resulting matrix is
\[
M' = IT_{d,0}^{-1} = {\small\begin{pmatrix} 1 & d & 0 \\ 0 & 1 & 0 \\ 0 & 0 & 1 \end{pmatrix}},
\]
which is exactly the expanded form of the propagation matrix in Table~\ref{tab:summary}.

In some applications (e.g.\ when mechanical dimensions are needed for prototyping an optomechanical layout) it is preferable to restore the original coordinate system.
This choice of output coordinates is accomplished by
\begin{equation}\label{eq:transform}
M' = T_{u,v} R_\theta M R_\theta^{-1} T_{u,v}^{-1}.
\end{equation}
This construction places the image space coordinate system to be coincident with the object space coordinate system,
which contrasts with the typical convention for ray transfer matrices that treats the object space and image space independently.

Several expressions other than Eq.~\eqref{eq:RTR} are found in the literature for handling decentered or tilted optical elements.\cite{gerrard1994,arnaud1976,shaomin1985,siegman1986,tovar1995,lin2006,lin2009,Lin2014}
For example, the rotation operator is often given as \cite{gerrard1994,siegman1986}
\[
R'_\theta = {\small\begin{pmatrix}
    1 & 0 & 0 \\
    0 & 1 & -\theta \\
    0 & 0 & 1
\end{pmatrix}}.    
\]
This operator is in fact not a rotation but a \emph{shear}, a vertical displacement proportional to the horizontal coordinate.
A shear is approximately equal to a rotation for sufficiently small angles, but we prefer the exact rotation formalism.

The benefit we gain over other implementations is that our Eq.~\eqref{eq:RTR} does not require small parameters, for example, $|\theta| \ll 1$ in the rotation matrix, because we are transforming the whole coordinate system rather than just the rays themselves.
In particular, plane mirrors of any orientation may be exactly modelled (see Sec.~\ref{sec:refl}), allowing the model to include, for example, path-folding mirrors along a beam.
So long as our rays do not stray too far from our transformed optical axis, the paraxial approximation is still valid even if the mechanical layout of the system is not along a single axis.
A similar method is described by Lin\cite{Lin2014} for transforming coordinates in non-paraxial ray tracing (e.g.~for implementation by computers), but Lin does not make the connection that these results apply to the paraxial case as well.

\subsection{Example: tilted window}
\begin{figure}
    \begin{center}
        \includegraphics[scale=1.0]{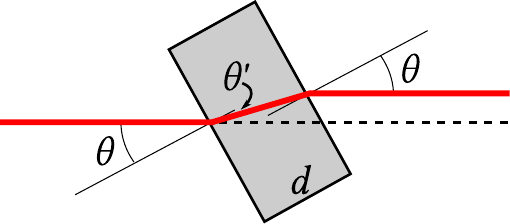}
        \caption{%
        (Color online)
        Tilted window of thickness $d$ and index of refraction $n$ at an angle $\theta$ relative to normal incidence.  The angle of incidence inside the window is denoted $\theta'$.
        The path of the ray is shown as a thick red line, with the original optical axis as a dashed line.}\label{fig:window}
    \end{center}
\end{figure}
Consider a window of thickness $d$ and index of refraction $n$ tilted by a small angle $\theta$ (Fig.~\ref{fig:window}).
The ray transfer matrix for this system referenced back to the incoming coordinate system is (remembering to read the order of terms from right to left)
\begin{align*}
M &= \underbrace{R_\theta T_{d,0}}_{\shortstack{\scriptsize return to\\ \scriptsize original coords.}}
\underbrace{M_\text{ref}(n,1)}_{\shortstack{\scriptsize back \\ \scriptsize surface}}
\underbrace{T_{d,0}^{-1}}_{\shortstack{\scriptsize window \\ \scriptsize thickness}}
\underbrace{M_\text{ref}(1,n)}_{\shortstack{\scriptsize front \\ \scriptsize surface}}
\underbrace{R_\theta^{-1}}_{\text{rotate}}, \\
&\approx {\small \begin{pmatrix}
1 & -d\left(1-\tfrac{1}{n}\right) & -d\theta\left(1-\tfrac{1}{n}\right) \\
0 & 1 & 0 \\
0 & 0 & 1
\end{pmatrix}},
\end{align*}
where $\theta^2$ and higher-orders terms have been discarded in keeping with the paraxial approximation (rays must stay near the transformed optical axis).
Applying this RTM to an input ray $(0,0,1)^T$ coincident with the input optical axis gives an output ray
\[ {\small
    \begin{pmatrix}
        1 & -d\left(1-\tfrac{1}{n}\right) & -d\theta\left(1-\tfrac{1}{n}\right) \\
        0 & 1 & 0 \\
        0 & 0 & 1
        \end{pmatrix}
        \begin{pmatrix} 0 \\ 0 \\ 1 \end{pmatrix}}
    = {\small
    \begin{pmatrix}
        -d\theta\left(1-\tfrac{1}{n} \right) \\ 0 \\ 1
    \end{pmatrix}},
\]
showing that the outgoing ray is parallel to the original optical axis and has been displaced upward by a distance $d\theta\left(1-\tfrac{1}{n} \right)$.
This agrees to within the paraxial limit with the direct calculation of the displacement of a ray parallel to the optical axis through a tilted plate using Snell's Law (e.g.~Problem 2-8 of Pedrotti\cite{Pedrotti-chapter}), which is
\[
\Delta h = \frac{d\sin(\theta-\theta')}{\cos(\theta')} \approx {d\theta}\left(1-\frac{1}{n}\right),
\]
where the angle of incidence of the ray inside the window, $\theta'$, is given by $\sin(\theta')=\sin(\theta)/n$, and we use the small-angle approximation in keeping with the paraxial limit needed for consistency with the refracting surfaces.

\section{Orientation and Reflections}\label{sec:refl}
Reflections require special attention if we wish to include orientation in our homogeneous representation.\cite{dorst2020,stolfi2014}
The convention of direction introduced in Section~\ref{sec:rtm_homo} can be summarized: for the homogeneous ray ${r}=(c,a,b)^T$ representing the line $ax+by+c=0$, the ray is oriented left-to-right if $b>0$ and right-to-left if $b<0$.
(This explains our choice of signs for the ray vectors in homogeneous coordinates.)
The case $b=0$ (vertical rays) does not occur in the paraxial limit, but we may still identify rays with $a>0$ as going down and rays with $a<0$ as going up.
The final case is $a=b=0$, representing the line infinitely far away encircling the plane.
For these ideal lines $c>0$ circulates counter-clockwise, and $c<0$ goes clockwise.
The null ray vector $(0,0,0)^T$ is undefined geometrically.

\begin{figure}
    \begin{center}
        \includegraphics{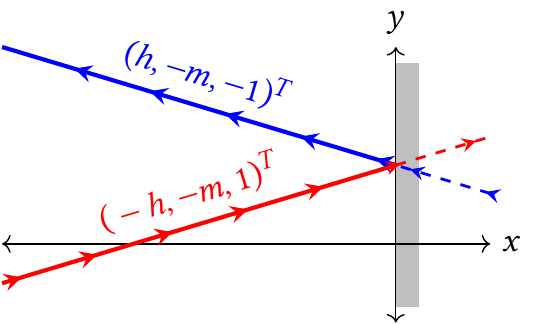}
        \caption{%
        (Color online)
        Incoming ray (red) striking a plane mirror and reflecting (blue). Both the ray's slope and its orientation change sign, resulting in the ray vector shown.}\label{fig:mirror}
    \end{center}
\end{figure}

Now consider the action of a plane mirror coincident with the $y$ axis on an incoming ray $(c,a,b)^T=(-h,-m,1)^T$ (Fig.~\ref{fig:mirror}).
We require both the slope of the ray and its orientation to change signs.
The RTM equation that satisfies this is
\[{\small
\begin{pmatrix} -1 & 0 & 0 \\ 0 & 1 & 0 \\ 0 & 0 & -1 \end{pmatrix}
\begin{pmatrix} -h \\ -m \\ 1 \end{pmatrix}}
= {\small
\begin{pmatrix} h \\ -m \\ -1 \end{pmatrix}}.
\]
The presence of the negative coefficient in the bottom right element of the RTM is the indicator of reflection.
Generalizing, we see that the recipe for converting the standard $2\times 2$ RTM of reflective elements into an RTM for oriented homogeneous coordinates is 
\[
M_\text{reflective} = 
(-1){\small \begin{pmatrix}
    A & B & 0 \\
    C & D & 0 \\
    0 & 0 & 1
\end{pmatrix}}.
\]
Specifically, the RTMs for a plane mirror normal to the axis and a spherical mirror with center of curvature on the axis are
\begin{align*}
    M_\text{plane mirror} &= 
    {\small
    \begin{pmatrix} -1 & 0 & 0 \\ 0 & 1 & 0 \\ 0 & 0 & -1 \end{pmatrix}},
    \\
    M_\text{sph.~mirror} &= 
    {\small
    \begin{pmatrix}
        -1 & 0 & 0 \\
        2/R & 1 & 0 \\
        0 & 0 & -1
    \end{pmatrix}},
\end{align*}
where $R>0$ for a convex mirror.

\subsection{Example: Retroreflector}
\begin{figure}
    \begin{center}
        \includegraphics{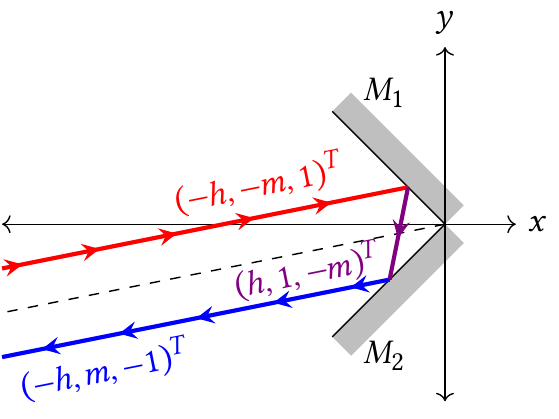}
        \caption{%
        (Color online)
        Incoming ray (red) striking two mutually perpendicular mirrors ($M_1$ and $M_2$) and reflecting twice (purple and blue rays).
        The outgoing ray is antiparallel to the incoming ray.
        The dashed line shows the bisector of the incoming and outgoing rays, which passes through the intersection of the mirrors.}\label{fig:retro}
    \end{center}
\end{figure}
A retroreflector can be made from a pair of plane mirrors intersecting at right angles (Fig.~\ref{fig:retro}).
Using our rotation operator from Eq.~\eqref{eq:RTR} we can build a system ray transfer matrix that also respects the orientation of the rays.
Note that our rotation angles are $\pm 45^{\circ}$, which are not small.
Care must be taken when ordering the RTMs.
We'll assume that the incoming ray ${r}_0 = (-h,-m,1)^T$ first strikes the upper mirror and then strikes the lower mirror.
Choosing a different input ray may require the user to change the order of the mirrors, depending on which surface is struck by the ray first.

The RTM for the upper mirror ($M_1$) is generated by a $45^{\circ}$ rotation of a plane mirror situated at the origin:
\begin{align}
M_1 &= R_{45^{\circ}}M_\text{plane mirror}R_{45^{\circ}}^{-1},\label{eq:retro} \\
&= {\small\begin{pmatrix}
    1 & 0 & 0 \\
    0 & \frac{1}{\sqrt{2}} & -\frac{1}{\sqrt{2}} \\
    0 & \frac{1}{\sqrt{2}} & \frac{1}{\sqrt{2}}
\end{pmatrix}
\begin{pmatrix} -1 & 0 & 0 \\ 0 & 1 & 0 \\ 0 & 0 & -1 \end{pmatrix}
\begin{pmatrix}
    1 & 0 & 0 \\
    0 & \frac{1}{\sqrt{2}} & \frac{1}{\sqrt{2}} \\
    0 & -\frac{1}{\sqrt{2}} & \frac{1}{\sqrt{2}}
\end{pmatrix}},\notag\\
&= {\small\begin{pmatrix}
-1 & 0 & 0 \\ 0 & 0 & 1 \\ 0 & 1 & 0
\end{pmatrix}}.\notag
\end{align}
Similarly, the lower mirror ($M_2$) has the RTM
\begin{equation}\label{eq:retroM2}
M_2 = R_{-45^{\circ}}M_\text{plane mirror}R_{-45^{\circ}}^{-1} = {\small\begin{pmatrix} -1 & 0 & 0 \\ 0 & 0 & -1 \\ 0 & -1 & 0 \end{pmatrix}}.    
\end{equation}
We do not need any translation operator between the two mirrors because both mirrors pass through the coodinate origin.
The passage of the ray from one mirror to the next is not explictly part of the RTM (other than the order of the optical elements).
In other words, the position of the optical elements determines the RTM, not the path of the rays.
We choose an incoming ray ${r}_0 = (-h,-m,1)^T$ with $h>0$ such that it will strike mirror $M_1$ first, yielding the reflected ray
\[
{r}_1 = M_1 {r}_0 = (h,1,-m)^T.    
\]
This ray follows the line $h +x-my =0$, propagating from right to left ($b<0$).
After the second reflection in $M_2$ the final ray is
\[
{r}_2 = M_2 {r}_1 = (-h,m,-1)^T,    
\]
which is antiparallel to the incoming ray, as expected, with a $y$ intercept of $-h$ and propagating right to left ($b<0$).
Note also that the method works for any angular separation between the mirrors by including the appropriate angles in Eqs.~\eqref{eq:retro} and \eqref{eq:retroM2}.

\section{Point transfer matrices}\label{sec:ptm}
Our next goal is to find a linear operator for our optical system that maps points in the object space of our optical system onto points in the image space.
We'll call this operator the \emph{point transfer matrix} (PTM) of the system in analogy to the ray transfer matrices discussed above.
Before we do that, we need to describe how we treat points.

\subsection{Homogeneous points}
We express a point with coordinates $[x,y]$ using \emph{homogeneous coordinates} as a column vector $[1,x,y]^T$.
(We use square brackets to denote point vectors and round brackets to denote ray vectors to avoid ambiguity.)
The added dimension will allow us to rotate and translate points similarly to how we manipulated rays above in Section~\ref{sec:refl}.
The term ``homogeneous'' means that we can multiply our point vector by any non-zero scalar without changing its physical meaning.
More generally, we express points as ${p} = [w, x, y]^T$.
We'll say the points are ``normalized'' if $w=1$.
For other $w\neq 0$, the corresponding physical point is located at position $[x/w,y/w]$.
For $w=0$, we interpret the point to be infinitely far away in the direction of slope $y/x$ (taking the limit $w\rightarrow 0^+$).
The inclusion of infinite (or ``ideal'') points is a key feature of homogeneous coordinates.
In optics we can use this notation to describe objects infinitely (or practically infinitely) far away, such as distant stars.
For additional details about the algebraic relationships between rays and points, please see Section 1 of the Supplemental Materials.
Generalizations to three dimensions (and more) can be found in Refs.~\onlinecite{stolfi2014,pharr2016,doran2007,winitzki,dorst2007,dorst2020}.

\subsection{Maintaining coincidence}
\begin{figure}
    \begin{center}
        \includegraphics{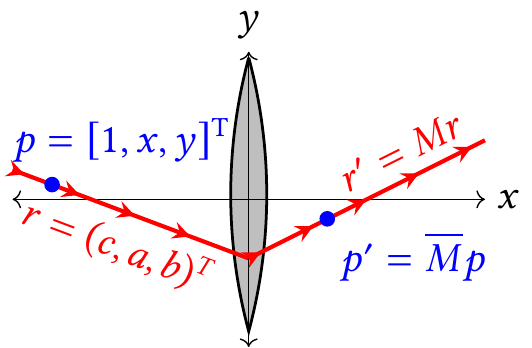}
        \caption{\label{fig:thinlens}%
        (Color online)
        Schematic representation of the effects of the ray transfer matrix $M$ and the point transfer matrix $\overline{M}$.
        The incoming ray $r$ (red line) intesects an object point $p$
        (blue point).
        The lens creates the outgoing ray $r'$ which must intersect the image point $p'$.
        The ray vectors are related by the ray transfer matrix by $r' = Mr$, and the point vectors are related by the corresponding point transfer matrix by $p'=\overline{M}p$.
        }
    \end{center}
\end{figure}
To motivate our definition for point transfer matrices, consider a ray ${r}=(c,a,b)^T$
that passes through a point $[x,y]$, represented by in homogeneous coordinates as the column vector ${p}= [1,x,y]^T$ (see Figure~\ref{fig:thinlens})
The coincidence relationship from above can be stated
\begin{equation*}
    c + ax + by = {p}^T {r}  =0.
\end{equation*}

Now, let's image this ray and point through an optical system with ray transfer matrix $M$ and yet-to-be-determined point transfer matrix $\overline{M}$:
\[
{r}' = M {r}, \qquad
{p}' = \overline{M} {p}.    
\]
The resulting image-space ray ${r}'$ and point ${p}'$ should also be coincident.
This requires
\begin{equation*}
    0={{p}'}^T {r}' =
    \left(\overline{M} {p} \right)^T \left( M {r} \right) =
    {p}^T (\overline{M}^T M) {r}.
\end{equation*}

The final expression reproduces the original coincidence relation provided that $\overline{M}^T M$ is a scalar times the identity matrix, or equivalently, that $\overline{M}^T$ is a scalar times $M^{-1}$.
A more rigorous derivation (see Section 2 of the Supplemental Materials\cite{SMNote}) shows that the proper scalar factor is $\det(M)$, yielding a point transfer matrix
\begin{equation}\label{eq:adjugate}
\overline{M} = \det(M)\, (M^{-1})^{T}.
\end{equation}
$\overline{M}$ can be calculated using this expression because the ray transfer matrix $M$ is necessarily invertible.
If the $2\times2$ RTM is known, the corresponding PTM is
\begin{equation}\label{eq:ptm}
    \overline{M}= {\small
\begin{bmatrix}
    D & -C & 0 \\ -B & A & 0 \\ 0 & 0 & (AD-BC)
\end{bmatrix}}.
\end{equation}
In the more general case (e.g.~when using the coordinate transformations of Sec.~\ref{sec:refl}), Eq.~\eqref{eq:adjugate} holds.

One can also show that $\overline{M}$ is equal to be the adjugate\cite{strang2006} of the ray transfer matrix $M$, which can be calculated even in the (unphysical) case $\det(M)=0$ using cofactor expansion.
When combining optical elements
\begin{align*}
\overline{(M_2M_1)}&=\det(M_2M_1)\bigl((M_2M_1)^{-1}\bigr)^T,\\
&= \det(M_2)\det(M_1)(M_1^{-1}M_2^{-1})^T, \\
&= \det(M_2)(M_2^{-1})^T\,\det(M_1)(M_1^{-1})^T
=\overline{M}_2\overline{M}_1,
\end{align*}
so the PTMs get multiplied in order from right to left, just like the RTMs.

In summary, if we know the ray transfer matrix $M$ of the optical system, which transforms rays using $r'=Mr$, then we can use Eq.~\eqref{eq:adjugate} to calculate the equivalent point transfer matrix $\overline{M}$ that transforms points using $p' = \overline{M}p$.

\subsection{Dimensional analysis and units}
As in the traditional ABCD formalism, our matrices and vectors contain mixed dimensionality.  For completeness, the units of the matrix elements of our RTM and PTM are given here, where $L$ is the unit of length.
\begin{align*}
RTM &\sim {\small\begin{pmatrix}
    1 & L & L \\
    L^{-1} & 1 & 1 \\
    L^{-1} & 1 & 1
\end{pmatrix}}, &
PTM &\sim {\small\begin{bmatrix}
    1 & L^{-1} & L^{-1} \\
    L & 1 & 1 \\
    L & 1 & 1
\end{bmatrix}}.
\end{align*}
The ray vectors $r$ and point vectors $p$ in normalized form have the dimensions:
\begin{align*}
    r &\sim (L, 1, 1)^T, & 
    p &\sim [1,L,L]^T.
\end{align*}
Because we are using homogeneous representations, the vectors may be multiplied by a non-zero dimensionful scalar without changing their geometric meaning.

\section{Point-transfer examples}\label{sec:ex}
Here we present four examples showing how the PTMs simplify calculations.

\subsection{Deriving Gauss's lens equation}
As a simple example employing the point-transfer matrix consider a thin lens of focal length $f$.
Reading the ray transfer matrix elements $ABCD$ from Table~\ref{tab:summary} and inserting them into the point transfer matrix, we get the equation
\begin{equation}\label{eq:thinlens}{\small
    \begin{bmatrix} w' \\ x' \\ y' \end{bmatrix}}
    = {\small
    \begin{bmatrix} 1 & 1/f & 0 \\ 0 & 1 & 0 \\ 0 & 0 & 1 \end{bmatrix}
    \begin{bmatrix} 1 \\ x \\ y \end{bmatrix}}
    = {\small
    \begin{bmatrix} 1+\frac{x}{f} \\ x \\ y \end{bmatrix}},
\end{equation}
where the input point is the normalized point at position $[x,y]$.
(Recall our sign convention for the coordinates as shown in Fig.~\ref{fig:ABCD}: $x$ and $x'$ increase along the direction of propagation, so that if we follow the usual convention in our optical drawings of light entering from the left, a real object has $x<0$ and a real image has $x'>0$.)
The resulting output is the normalized point
\begin{align*}
\hat{x}' &= \frac{x'}{w'} = \frac{x}{1+(x/f)} = \frac{1}{(1/x)+(1/f)},\\
\hat{y}' &= \frac{y'}{w'} = \frac{y}{1+(x/f)} = \frac{\hat{x}'}{x}\,y.
\end{align*}
The first line agrees with the Gaussian lens equation giving the location of the image, and the second line gives the image magnification $\hat{x}'/x$.
The sign of $w'$ before normalization gives the orientation of the image: $w'>0$ indicates upright and $w'<0$ indicates inverted.
Our formulation also easily handles the special case when the object (image) point is infinitely far away (the homogeneous weight $w$ ($w'$) goes to zero) as well as virtual objects and virtual images, corresponding respectively to  $x/w>0$ and $x'/w'<0$ in the usual case where the optical system is aligned with the $x$-axis.

\subsection{Imaging an infinite object point}
To see how the PTMs are used in numerical calculations, consider a simple single-lens camera pointed at the horizon to photograph a star infinitely far away.
The parallel rays from the star will converge at the back focal plane of the lens, at a height equal to minus the angular height of the star, times the focal length of the lens.

\begin{figure}
    \begin{center}
        \includegraphics{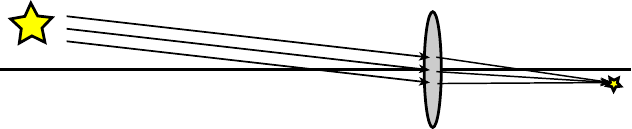}
        \caption{%
        (Color online)
        Calculating the image of a star by treating the star as an object point infinitely far away.
        The location of the star in homogeneous coordinates is $[0,-1,0.01]^T$, the lens has a focal length of $f={50}\,\mathrm{mm}$, and the star's image is at $[1,x,y]^T=[1,50,-0.5]^T$ (after normalization).
        The length units of the vectors are millimeters.
        }\label{fig:star}
    \end{center}
\end{figure}

For concreteness, we can also show this using our point transfer matrix for a star $10$ milliradians above the horizon (Fig.~\ref{fig:star}) and an $f={50}\,\mathrm{mm}$ lens.
The resulting image of the star may be found using the point transfer matrix for a thin lens:
\[ {\small
\begin{bmatrix}
1 & 1/50 & 0 \\
0 & 1 & 0 \\
0 & 0 & 1
\end{bmatrix}
\begin{bmatrix} 0 \\ -1 \\ 0.01 \end{bmatrix}}
= {\small
\begin{bmatrix} -0.02 \\ -1 \\ 0.01\end{bmatrix}}
\equiv {\small
\begin{bmatrix} 1 \\ 50 \\ -0.5\end{bmatrix}},
\]
with implied length units of millimeters.
The last step is normalization of the point vector.
In this example, the object point is infinitely far away in the direction of slope $= \arctan(0.01/-1)\approx -0.01$ (so the star is above the optical axis ($y>0$) and before the lens ($x<0$)).
The final normalized image point has height $y'={-0.5}\,\mathrm{mm}$ and is located at $x'={50}\,\mathrm{mm}$ after the lens, corresponding with the back focal plane of the lens as expected.
Note that we do not need any explicit reference to the light rays to perform the calculation.
This example demonstrates one advantage of this method: using homogeneous coordinates allows us to treat both finite and infinite objects/images in the same way without resorting to special cases.
A naïve application of Gauss's lens equation with an infinite object distance would not tell us the vertical position of the image because the linear magnification is undefined in this instance.

\subsection{Analysis of a compound lens}
For our next example (adapted from Examples 6.6 and 6.7 of Hecht\cite{Hecht-chapter} -- note that we use a different convention for the ray transfer matrices than Hecht), consider a compound lens system with the ray transfer matrix (length units of cm)
\[
M=\begin{pmatrix} A & B \\ C & D \end{pmatrix}
=
\begin{pmatrix}
    0.867 & 1.338 \\ -0.198 & 0.848
  \end{pmatrix},
\]
and an object 20 cm in front of the lens with a height of 0.1 cm.
We would like to locate the resulting image position and height.

Hecht's solution has several steps: 
(1) Extend the system matrix $M$ with a propagation matrix of length 20 cm in front of the lens (for the object position) and a propagation of unknown length $d$ after the lens (for the image position) so that the system matrix now includes both the object and image points.
(2) Algebraically set the $B$ element of the resulting product matrix to zero.  This enforces the condition that the final ray height is independent of the incoming ray slope (i.e.\ an image forms). With this constraint, one can solve for the unknown length $d$ to locate the image plane.
(3) Identify the magnification as the $A$ element of the resulting matrix.  This is true because we set $B=0$ above, meaning that the image ray height only depends on the object ray height, which gives us the magnification.

Our solution is much more direct.
First, we construct the point transfer matrix from the original ray transfer matrix (without additional propagations) using Eq.~\ref{eq:ptm} and multiply this into the homogeneous position vector $p=[1,x,y]^T$ for the object point.
This gives us the image point $p'$.
\begin{align*} 
  p'=
    \overline{M} p &=
    {\small
      \begin{bmatrix}
        0.848 & 0.198 & 0 \\
        -1.338 & 0.867 & 0 \\
        0 & 0 & 1
      \end{bmatrix}
      \begin{bmatrix}
        1 \\ -20.0 \\ 0.1
      \end{bmatrix}},\\
      &= {\small
      \begin{bmatrix}
        -3.112 \\ -18.678 \\ 0.1
      \end{bmatrix}}
      \equiv {\small
      \begin{bmatrix}
        1 \\ 6.002 \\ -0.032
      \end{bmatrix}}.
    \end{align*}
After performing the matrix multiplication, we only need to normalize the resulting image point vector (final step above) to read off the coordinates of the image: ${6.002}\,\mathrm{cm}$ past the lens at a height of ${-0.032}\,\mathrm{cm}$.
We've been able to replace algebraic multiplication of matrices containing unknown variables and solving for an unknown distance with a direct numerical calculation.

Taking this example one step farther, we can also calculate the front and back focal points of the lens system easily.
The back focal point (BFP) is the image of an ideal (infinite) object point on the axis: $[0,-1,0]^T$.  The image of this point is
\[ {\small
    \begin{bmatrix}
        0.848 & 0.198 & 0 \\
        -1.338 & 0.867 & 0 \\
        0 & 0 & 1
      \end{bmatrix}
      \begin{bmatrix} 0 \\ -1 \\ 0 \end{bmatrix}}
      = 
      {\small
      \begin{bmatrix}-0.198 \\ -0.867 \\ 0 \end{bmatrix}}
      \equiv
      {\small
      \begin{bmatrix} 1 \\ 4.38 \\ 0 \end{bmatrix}},
\]
where the last step is normalization.
Reading off the coordinates, the BFP is located 4.38 cm past the final optical surface.
To find the front focal point (FFP), we invert the point-transfer matrix of the system (i.e. time reversal) and then follow the same procedure using a ray coming from the reverse direction, $[0,1,0]^T$.

\subsection{Misaligned thin lens}
A common laboratory task is to take a collimated laser beam and focus it through a pinhole located at the back focal point of a converging lens (for example, to build a spatial filter or to inject the beam into a fiber).
If the pinhole is in a fixed position, but the beam doesn't quite hit the hole, a common remedy is to move the lens slightly in the transverse direction.
Another common tactic in the lab is to intentionally tilt a lens to prevent back-reflections from causing interference fringes.
As long as these shifts are small, the induced aberrations can be neglected.
Here we analyze this configuration for small displacements and small tilts of the lens.

Let's model our laser beam as a point source infinitely far away, parallel to the optical axis, and ignore diffraction effects.
The corresponding (infinite) point vector is ${p}=[0,1,0]^T$.
In essence, we're approximating the laser beam as a pencil of parallel rays.
We'll assume a converging lens of focal length $f$.
First, let's consider what happens to the image if we translate the lens by a distance $d$ perpendicular to the optical axis.
We do this by applying the point translation operator $\overline{T}_{0,d}$ to the lens PTM.
The resulting system matrix $\overline{M}$ is
\begin{align*}
\overline{M} &= \overline{T}_{0,d}\, \overline{M}_\text{thin lens}(f)\, \overline{T}^{-1}_{0,d},\\
  &=
  {\small
  \begin{bmatrix} 
    1 & 0 & 0 \\
    0 & 1 & 0 \\
    d & 0 & 1
    \end{bmatrix}
    \begin{bmatrix}
        1 & 1/f & 0 \\
        0 & 1 & 0 \\
        0 & 0 & 1
    \end{bmatrix}
    \begin{bmatrix} 
        1 & 0 & 0 \\
        0 & 1 & 0 \\
        -d & 0 & 1
        \end{bmatrix}},\\
  &= {\small
  \begin{bmatrix}
    1 & 1/f & 0 \\
    0 & 1 & 0 \\
    0 & d/f & 1
\end{bmatrix}},
\end{align*}
and the image point ${p}'$ is
\[
{p}'=\overline{M}{p}=
{\small
\begin{bmatrix}
    1 & 1/f & 0 \\
    0 & 1 & 0 \\
    0 & d/f & 1
\end{bmatrix}
\begin{bmatrix}
0 \\ 1 \\ 0 
\end{bmatrix}}
= {\small
\begin{bmatrix}
1/f \\ 1 \\ d/f
\end{bmatrix}}
\equiv
{\small
\begin{bmatrix}
    1 \\ f \\ d
\end{bmatrix}},
\]
where the last step is normalization of the point vector.
We see that the beam is still focused on the back focal plane ($x'=f$), but the focus point is displaced vertically by the same amount as the lens, $d$.

If we tilt the lens, the system PTM is
\begin{align*}
\overline{M} &= \overline{R}_{\theta}\, \overline{M}_\text{thin lens}(f)\, \overline{R}^{-1}_{\theta},\\
  &= {\small
  \begin{bmatrix} 
    1 & 0 & 0 \\
    0 & \cos\theta & -\sin\theta \\
    0 & \sin\theta & \cos\theta
    \end{bmatrix}
    \begin{bmatrix}
        1 & 1/f & 0 \\
        0 & 1 & 0 \\
        0 & 0 & 1
    \end{bmatrix}
    \begin{bmatrix} 
        1 & 0 & 0 \\
        0 & \cos\theta & \sin\theta \\
        0 & -\sin\theta & \cos\theta
            \end{bmatrix}},\\
  &=  
  {\small
  \begin{bmatrix}
    1 & \cos(\theta)/f & \sin(\theta)/f \\
    0 & 1 & 0 \\ 
    0 & 0 & 1
\end{bmatrix}}.
\end{align*}
Applying this to the object point, we get
\begin{align*}
{p}' = \overline{M}{p} &= {\small
\begin{bmatrix}
    1 & \cos(\theta)/f & \sin(\theta)/f \\
    0 & 1 & 0 \\ 
    0 & 0 & 1
\end{bmatrix}
\begin{bmatrix} 0 \\ 1 \\ 0 \end{bmatrix}}, \\
&= {\small
\begin{bmatrix} \cos(\theta)/f \\ 1 \\ 0 \end{bmatrix}}
\equiv {\small
\begin{bmatrix} 1 \\ f/\cos(\theta) \\ 0 \end{bmatrix}},\\
&\approx {\small
\begin{bmatrix} 1 \\ f (1+\theta^2/2) \\ 0 \end{bmatrix}}
.
\end{align*}
So, up to first order in the angle $\theta$, tilting the lens has no effect on the location of the beam focus.

\section{Outlook and Conclusion}\label{sec:3d}
The results we've presented (summarized in Table \ref{tab:summary}) are restricted to two-dimensional systems or three-dimensional systems with axial symmetry.
To consider general three-dimensional systems, we will need a different approach.
The homogeneous ray transfer matrices are linearizations of the propagation equations of Hamiltonian optics,\cite{Born1999,wolf2004} restricted to the paraxial regime.
A similar approach can be used to expand the work here to three-dimensional optics.
Rays in three dimensions may be represented homogeneously in 3D using a set of 6 Plüker coordinates.\cite{plueker1865,stolfi2014,lin2009}
These may be further reduced to 5 non-trivial coordinates for paraxial rays in 3D (e.g.~near the $z$ axis).\cite{lin2006}
Geometric Algebra,\cite{hestenes2003,macdonald2011,doran1993,doran2007,dorst2007,dorst2020} following Dorst's description of the algebra of lines\cite{dorst2016} and earlier work on 2D optical systems by Sugon and McNamara,\cite{sugon2003,sugon2006,sugon2008a,sugon2008b} is an intriguing mathematical system for approaching this problem that we are looking into.

Although the path to describing rays in 3D is straight-forward, the key difficulty in extending the point transfer matrices to 3D is that points in 3D do not necessarily image onto points.
A converging cylindrical lens is a simple example: it images a point source of light onto a line.
The general behavior in 3D requires defining a point as the intersection of three non-coplanar lines (not two lines, as one might naïvely assume),\cite{dorst2016} and then studying how those lines transform, noting that the three output lines may not converge to the same image point (e.g.~because of astigmatism).
This leads to the geometric theory of line complexes.\cite{pottman2001}
As a tantalizing bonus, Arnaud's representation of Gaussian laser beams as complex-valued rays\cite{arnaud1976,arnaud1985} may also be interpreted as skew line complexes.\cite{colbourne2014}
We are still investigating how to use this correspondence to build a geometric representation of the paraxial optics of Gaussian beams that is more flexible than the complex beam parameter formalism.\cite{siegman1986} The complex beam parameter itself comes from the  expression of diffraction integrals using RTMs described by Collins\cite{collins1970} and the related extension of the diffraction integrals by Bandres to include small rotations and translations.\cite{bandres2009}

In conclusion, we have shown that enhancing the ABCD ray transfer matrices (RTM) by applying them in a homogeneous coordinate system greatly expands their usefulness.
With a fairly modest change in interpretation of the RTM to better capture their geometric content we added the capability to describe translated and rotated optical elements and keep track of the ray propagation directions.
Lastly, we demonstrated that a simple mathematical operation converts the RTMs into point transfer matrices (PTM), simplifying imaging calculations.
We hope that our examples demonstrate the advantages of using this system and can be a jumping-off point for more advanced applications.

\begin{acknowledgments}
We gratefully acknowledge Christian Faber, Stephen De Keninck, and Leo Dorst for helpful suggestions, especially regarding the underlying geometric ideas here.
\end{acknowledgments}

\bibliography{paper} 

\begin{thebibliography}{42}%
\makeatletter
\providecommand \@ifxundefined [1]{%
 \@ifx{#1\undefined}
}%
\providecommand \@ifnum [1]{%
 \ifnum #1\expandafter \@firstoftwo
 \else \expandafter \@secondoftwo
 \fi
}%
\providecommand \@ifx [1]{%
 \ifx #1\expandafter \@firstoftwo
 \else \expandafter \@secondoftwo
 \fi
}%
\providecommand \natexlab [1]{#1}%
\providecommand \enquote  [1]{``#1''}%
\providecommand \bibnamefont  [1]{#1}%
\providecommand \bibfnamefont [1]{#1}%
\providecommand \citenamefont [1]{#1}%
\providecommand \href@noop [0]{\@secondoftwo}%
\providecommand \href [0]{\begingroup \@sanitize@url \@href}%
\providecommand \@href[1]{\@@startlink{#1}\@@href}%
\providecommand \@@href[1]{\endgroup#1\@@endlink}%
\providecommand \@sanitize@url [0]{\catcode `\\12\catcode `\$12\catcode
  `\&12\catcode `\#12\catcode `\^12\catcode `\_12\catcode `\%12\relax}%
\providecommand \@@startlink[1]{}%
\providecommand \@@endlink[0]{}%
\providecommand \url  [0]{\begingroup\@sanitize@url \@url }%
\providecommand \@url [1]{\endgroup\@href {#1}{\urlprefix }}%
\providecommand \urlprefix  [0]{URL }%
\providecommand \Eprint [0]{\href }%
\providecommand \doibase [0]{http://dx.doi.org/}%
\providecommand \selectlanguage [0]{\@gobble}%
\providecommand \bibinfo  [0]{\@secondoftwo}%
\providecommand \bibfield  [0]{\@secondoftwo}%
\providecommand \translation [1]{[#1]}%
\providecommand \BibitemOpen [0]{}%
\providecommand \bibitemStop [0]{}%
\providecommand \bibitemNoStop [0]{.\EOS\space}%
\providecommand \EOS [0]{\spacefactor3000\relax}%
\providecommand \BibitemShut  [1]{\csname bibitem#1\endcsname}%
\let\auto@bib@innerbib\@empty
\bibitem [{\citenamefont {Halbach}(1964)}]{halbach1964}%
  \BibitemOpen
  \bibfield  {author} {\bibinfo {author} {\bibfnamefont {Klaus}\ \bibnamefont
  {Halbach}},\ }\bibfield  {title} {\enquote {\bibinfo {title} {Matrix
  representation of {{Gaussian}} optics},}\ }\href {\doibase 10.1119/1.1970159}
  {\bibfield  {journal} {\bibinfo  {journal} {American Journal of Physics}\
  }\textbf {\bibinfo {volume} {32}},\ \bibinfo {pages} {90--108} (\bibinfo
  {year} {1964})}\BibitemShut {NoStop}%
\bibitem [{\citenamefont {Gerrard}\ and\ \citenamefont
  {Burch}(1994)}]{gerrard1994}%
  \BibitemOpen
  \bibfield  {author} {\bibinfo {author} {\bibfnamefont {A.}~\bibnamefont
  {Gerrard}}\ and\ \bibinfo {author} {\bibfnamefont {J.~M.}\ \bibnamefont
  {Burch}},\ }\href@noop {} {\emph {\bibinfo {title} {Introduction to Matrix
  Methods in Optics}}}\ (\bibinfo  {publisher} {{Dover}},\ \bibinfo {address}
  {{New York}},\ \bibinfo {year} {1994})\BibitemShut {NoStop}%
\bibitem [{\citenamefont {Pedrotti}\ \emph {et~al.}(2007)\citenamefont
  {Pedrotti}, \citenamefont {Pedrotti},\ and\ \citenamefont
  {Pedrotti}}]{Pedrotti-chapter}%
  \BibitemOpen
  \bibfield  {author} {\bibinfo {author} {\bibfnamefont {Frank~L.}\
  \bibnamefont {Pedrotti}}, \bibinfo {author} {\bibfnamefont {Leno~M.}\
  \bibnamefont {Pedrotti}}, \ and\ \bibinfo {author} {\bibfnamefont {Leno~S.}\
  \bibnamefont {Pedrotti}},\ }\href@noop {} {\emph {\bibinfo {title}
  {Introduction to Optics}}},\ \bibinfo {edition} {3rd}\ ed.\ (\bibinfo
  {publisher} {Pearson},\ \bibinfo {address} {Harlow},\ \bibinfo {year}
  {2007})\ Chap.~\bibinfo {chapter} {18}\BibitemShut {NoStop}%
\bibitem [{\citenamefont {Hecht}(2017)}]{Hecht-chapter}%
  \BibitemOpen
  \bibfield  {author} {\bibinfo {author} {\bibfnamefont {Eugene}\ \bibnamefont
  {Hecht}},\ }\href@noop {} {\emph {\bibinfo {title} {Optics}}},\ \bibinfo
  {edition} {5th}\ ed.\ (\bibinfo  {publisher} {Pearson},\ \bibinfo {year}
  {2017})\ Chap.~\bibinfo {chapter} {6}\BibitemShut {NoStop}%
\bibitem [{\citenamefont {Born}\ and\ \citenamefont {Wolf}(1999)}]{Born1999}%
  \BibitemOpen
  \bibfield  {author} {\bibinfo {author} {\bibfnamefont {Max}\ \bibnamefont
  {Born}}\ and\ \bibinfo {author} {\bibfnamefont {Emil}\ \bibnamefont {Wolf}},\
  }\href@noop {} {\emph {\bibinfo {title} {Principles of Optics}}},\ \bibinfo
  {edition} {7th}\ ed.\ (\bibinfo  {publisher} {Cambridge},\ \bibinfo {address}
  {New York},\ \bibinfo {year} {1999})\ Chap.~\bibinfo {chapter}
  {4}\BibitemShut {NoStop}%
\bibitem [{\citenamefont {Pharr}\ \emph {et~al.}(2016)\citenamefont {Pharr},
  \citenamefont {Jakob},\ and\ \citenamefont {Humphreys}}]{pharr2016}%
  \BibitemOpen
  \bibfield  {author} {\bibinfo {author} {\bibfnamefont {Matt}\ \bibnamefont
  {Pharr}}, \bibinfo {author} {\bibfnamefont {Wenzel}\ \bibnamefont {Jakob}}, \
  and\ \bibinfo {author} {\bibfnamefont {Greg}\ \bibnamefont {Humphreys}},\
  }\href {http://pbrt-book.org} {\emph {\bibinfo {title} {Physically Based
  Rendering: From Theory to Implementation}}},\ \bibinfo {edition} {3rd}\ ed.\
  (\bibinfo  {publisher} {{Morgan Kaufmann}},\ \bibinfo {address} {{Cambridge,
  MA}},\ \bibinfo {year} {2016})\BibitemShut {NoStop}%
\bibitem [{\citenamefont {Stolfi}(2014)}]{stolfi2014}%
  \BibitemOpen
  \bibfield  {author} {\bibinfo {author} {\bibfnamefont {Jorge}\ \bibnamefont
  {Stolfi}},\ }\href@noop {} {\emph {\bibinfo {title} {Oriented Projective
  Geometry: A Framework for Geometric Computations}}}\ (\bibinfo  {publisher}
  {{Academic Press}},\ \bibinfo {year} {2014})\BibitemShut {NoStop}%
\bibitem [{\citenamefont {Coxeter}(2003)}]{coxeter2003}%
  \BibitemOpen
  \bibfield  {author} {\bibinfo {author} {\bibfnamefont {H.~S.~M.}\
  \bibnamefont {Coxeter}},\ }\href@noop {} {\emph {\bibinfo {title}
  {{Projective Geometry}}}},\ \bibinfo {edition} {2nd}\ ed.\ (\bibinfo
  {publisher} {{Springer}},\ \bibinfo {address} {{New York}},\ \bibinfo {year}
  {2003})\BibitemShut {NoStop}%
\bibitem [{\citenamefont {Cambi}(1959)}]{cambi1959}%
  \BibitemOpen
  \bibfield  {author} {\bibinfo {author} {\bibfnamefont {Enzo}\ \bibnamefont
  {Cambi}},\ }\bibfield  {title} {\enquote {\bibinfo {title} {Projective
  formulation of the problems of geometrical optics. {I.} {Theoretical}
  foundations},}\ }\href {\doibase 10.1364/JOSA.49.000002} {\bibfield
  {journal} {\bibinfo  {journal} {Journal of the Optical Society of America}\
  }\textbf {\bibinfo {volume} {49}},\ \bibinfo {pages} {2--15} (\bibinfo {year}
  {1959})}\BibitemShut {NoStop}%
\bibitem [{\citenamefont {Feder}(1963)}]{feder1963}%
  \BibitemOpen
  \bibfield  {author} {\bibinfo {author} {\bibfnamefont {Donald~P.}\
  \bibnamefont {Feder}},\ }\bibfield  {title} {\enquote {\bibinfo {title}
  {Automatic optical design},}\ }\href {\doibase 10.1364/AO.2.001209}
  {\bibfield  {journal} {\bibinfo  {journal} {Applied Optics}\ }\textbf
  {\bibinfo {volume} {2}},\ \bibinfo {pages} {1209} (\bibinfo {year}
  {1963})}\BibitemShut {NoStop}%
\bibitem [{\citenamefont {Wynne}\ and\ \citenamefont
  {Wormell}(1963)}]{wynne1963}%
  \BibitemOpen
  \bibfield  {author} {\bibinfo {author} {\bibfnamefont {C.~G.}\ \bibnamefont
  {Wynne}}\ and\ \bibinfo {author} {\bibfnamefont {P.~M. J.~H.}\ \bibnamefont
  {Wormell}},\ }\bibfield  {title} {\enquote {\bibinfo {title} {Lens design by
  computer},}\ }\href {\doibase 10.1364/AO.2.001233} {\bibfield  {journal}
  {\bibinfo  {journal} {Applied Optics}\ }\textbf {\bibinfo {volume} {2}},\
  \bibinfo {pages} {1233--1238} (\bibinfo {year} {1963})}\BibitemShut {NoStop}%
\bibitem [{\citenamefont {Arnaud}(1976)}]{arnaud1976}%
  \BibitemOpen
  \bibfield  {author} {\bibinfo {author} {\bibfnamefont {J.~A.}\ \bibnamefont
  {Arnaud}},\ }\href@noop {} {\emph {\bibinfo {title} {Beam And Fiber
  Optics}}}\ (\bibinfo  {publisher} {{Academic}},\ \bibinfo {address} {{New
  York}},\ \bibinfo {year} {1976})\BibitemShut {NoStop}%
\bibitem [{\citenamefont {Shaomin}(1985)}]{shaomin1985}%
  \BibitemOpen
  \bibfield  {author} {\bibinfo {author} {\bibfnamefont {Wang}\ \bibnamefont
  {Shaomin}},\ }\bibfield  {title} {\enquote {\bibinfo {title} {Matrix methods
  in treating decentred optical systems},}\ }\href {\doibase 10/d79hb8}
  {\bibfield  {journal} {\bibinfo  {journal} {Optical and Quantum Electronics}\
  }\textbf {\bibinfo {volume} {17}},\ \bibinfo {pages} {1--14} (\bibinfo {year}
  {1985})}\BibitemShut {NoStop}%
\bibitem [{\citenamefont {Siegman}(1986)}]{siegman1986}%
  \BibitemOpen
  \bibfield  {author} {\bibinfo {author} {\bibfnamefont {A.~E.}\ \bibnamefont
  {Siegman}},\ }\href@noop {} {\emph {\bibinfo {title} {Lasers}}}\ (\bibinfo
  {publisher} {{University Science Books}},\ \bibinfo {year}
  {1986})\BibitemShut {NoStop}%
\bibitem [{\citenamefont {Tovar}\ and\ \citenamefont
  {Casperson}(1995)}]{tovar1995}%
  \BibitemOpen
  \bibfield  {author} {\bibinfo {author} {\bibfnamefont {Anthony~A.}\
  \bibnamefont {Tovar}}\ and\ \bibinfo {author} {\bibfnamefont {Lee~W.}\
  \bibnamefont {Casperson}},\ }\bibfield  {title} {\enquote {\bibinfo {title}
  {Generalized beam matrices: {{Gaussian}} beam propagation in misaligned
  complex optical systems},}\ }\href {\doibase 10.1364/JOSAA.12.001522}
  {\bibfield  {journal} {\bibinfo  {journal} {Journal of the Optical Society of
  America A}\ }\textbf {\bibinfo {volume} {12}},\ \bibinfo {pages} {1522--1533}
  (\bibinfo {year} {1995})}\BibitemShut {NoStop}%
\bibitem [{\citenamefont {Lin}\ and\ \citenamefont {Sung}(2006)}]{lin2006}%
  \BibitemOpen
  \bibfield  {author} {\bibinfo {author} {\bibfnamefont {Psang~Dain}\
  \bibnamefont {Lin}}\ and\ \bibinfo {author} {\bibfnamefont {Chi-Kuen}\
  \bibnamefont {Sung}},\ }\bibfield  {title} {\enquote {\bibinfo {title}
  {Matrix-based paraxial skew ray-tracing in {{3D}} systems with non-coplanar
  optical axis},}\ }\href {\doibase 10.1016/j.ijleo.2005.10.004} {\bibfield
  {journal} {\bibinfo  {journal} {Optik}\ }\textbf {\bibinfo {volume} {117}},\
  \bibinfo {pages} {329--340} (\bibinfo {year} {2006})}\BibitemShut {NoStop}%
\bibitem [{\citenamefont {Lin}\ and\ \citenamefont {Hsueh}(2009)}]{lin2009}%
  \BibitemOpen
  \bibfield  {author} {\bibinfo {author} {\bibfnamefont {P.~D.}\ \bibnamefont
  {Lin}}\ and\ \bibinfo {author} {\bibfnamefont {C.-C.}\ \bibnamefont
  {Hsueh}},\ }\bibfield  {title} {\enquote {\bibinfo {title} {6$\times$6 matrix
  formalism of optical elements for modeling and analyzing {{3D}} optical
  systems},}\ }\href {\doibase 10.1007/s00340-009-3616-7} {\bibfield  {journal}
  {\bibinfo  {journal} {Applied Physics B}\ }\textbf {\bibinfo {volume} {97}},\
  \bibinfo {pages} {135--143} (\bibinfo {year} {2009})}\BibitemShut {NoStop}%
\bibitem [{\citenamefont {Lin}(2014)}]{Lin2014}%
  \BibitemOpen
  \bibfield  {author} {\bibinfo {author} {\bibfnamefont {Psang~Dain}\
  \bibnamefont {Lin}},\ }\href@noop {} {\emph {\bibinfo {title} {New
  Computation Methods for Geometrical Optics}}},\ Vol.\ \bibinfo {volume}
  {178}\ (\bibinfo  {publisher} {Springer},\ \bibinfo {address} {Singapore},\
  \bibinfo {year} {2014})\BibitemShut {NoStop}%
\bibitem [{\citenamefont {Liu}\ and\ \citenamefont {Brenner}(2008)}]{liu2008}%
  \BibitemOpen
  \bibfield  {author} {\bibinfo {author} {\bibfnamefont {Xiyuan}\ \bibnamefont
  {Liu}}\ and\ \bibinfo {author} {\bibfnamefont {Karl-Heinz}\ \bibnamefont
  {Brenner}},\ }\bibfield  {title} {\enquote {\bibinfo {title} {Minimal optical
  decomposition of ray transfer matrices},}\ }\href {\doibase
  10.1364/ao.47.000e88} {\bibfield  {journal} {\bibinfo  {journal} {Applied
  Optics}\ }\textbf {\bibinfo {volume} {47}},\ \bibinfo {pages} {E88} (\bibinfo
  {year} {2008})}\BibitemShut {NoStop}%
\bibitem [{\citenamefont {Tovar}\ and\ \citenamefont
  {Casperson}(1997)}]{tovar1997}%
  \BibitemOpen
  \bibfield  {author} {\bibinfo {author} {\bibfnamefont {Anthony~A.}\
  \bibnamefont {Tovar}}\ and\ \bibinfo {author} {\bibfnamefont {Lee~W.}\
  \bibnamefont {Casperson}},\ }\bibfield  {title} {\enquote {\bibinfo {title}
  {Generalized beam matrices. {{IV}}. {{Optical}} system design},}\ }\href
  {\doibase 10.1364/JOSAA.14.000882} {\bibfield  {journal} {\bibinfo  {journal}
  {Journal of the Optical Society of America A}\ }\textbf {\bibinfo {volume}
  {14}},\ \bibinfo {pages} {882--894} (\bibinfo {year} {1997})}\BibitemShut
  {NoStop}%
\bibitem [{SMN()}]{SMNote}%
  \BibitemOpen
  \href@noop {} {}\bibinfo {note} {{Supplemental material is available at [url
  to be inserted by AIPP].}}\BibitemShut {Stop}%
\bibitem [{not()}]{note:atan}%
  \BibitemOpen
  \href@noop {} {}\bibinfo {note} {Equivalently, one may use the two-argument
  form of the arctangent: \(\phi = \mathrm{arctan2}(a,-b)\).}\BibitemShut
  {Stop}%
\bibitem [{\citenamefont {Dorst}(2020)}]{dorst2020}%
  \BibitemOpen
  \bibfield  {author} {\bibinfo {author} {\bibfnamefont {Leo}\ \bibnamefont
  {Dorst}},\ }\href {http://bivector.net/PGA4CS.html} {\enquote {\bibinfo
  {title} {A guided tour to the plane-based geometric algebra {PGA}},}\ }
  (\bibinfo {year} {2020}),\ \bibinfo {note} {intended as replacement for
  Ch.~11 of \cite{dorst2007}}\BibitemShut {NoStop}%
\bibitem [{\citenamefont {Doran}\ and\ \citenamefont
  {Lasenby}(2007)}]{doran2007}%
  \BibitemOpen
  \bibfield  {author} {\bibinfo {author} {\bibfnamefont {Chris}\ \bibnamefont
  {Doran}}\ and\ \bibinfo {author} {\bibfnamefont {A.~N.}\ \bibnamefont
  {Lasenby}},\ }\href@noop {} {\emph {\bibinfo {title} {Geometric Algebra for
  Physicists}}}\ (\bibinfo  {publisher} {{Cambridge University Press}},\
  \bibinfo {address} {{Cambridge}},\ \bibinfo {year} {2007})\BibitemShut
  {NoStop}%
\bibitem [{\citenamefont {Winitzki}(2020)}]{winitzki}%
  \BibitemOpen
  \bibfield  {author} {\bibinfo {author} {\bibfnamefont {Sergei}\ \bibnamefont
  {Winitzki}},\ }\href {https://github.com/winitzki/linear-algebra-book} {\emph
  {\bibinfo {title} {Linear Algebra via Exterior Products}}},\ \bibinfo
  {edition} {v. 1.3}\ ed.\ (\bibinfo  {publisher} {Lulu},\ \bibinfo {year}
  {2020})\BibitemShut {NoStop}%
\bibitem [{\citenamefont {Dorst}\ \emph {et~al.}(2007)\citenamefont {Dorst},
  \citenamefont {Fontijne},\ and\ \citenamefont {Mann}}]{dorst2007}%
  \BibitemOpen
  \bibfield  {author} {\bibinfo {author} {\bibfnamefont {Leo}\ \bibnamefont
  {Dorst}}, \bibinfo {author} {\bibfnamefont {Daniel}\ \bibnamefont
  {Fontijne}}, \ and\ \bibinfo {author} {\bibfnamefont {Stephen}\ \bibnamefont
  {Mann}},\ }\href@noop {} {\emph {\bibinfo {title} {Geometric Algebra for
  Computer Science}}}\ (\bibinfo  {publisher} {Morgan Kaufmann},\ \bibinfo
  {address} {Amsterdam},\ \bibinfo {year} {2007})\BibitemShut {NoStop}%
\bibitem [{\citenamefont {Strang}(2006)}]{strang2006}%
  \BibitemOpen
  \bibfield  {author} {\bibinfo {author} {\bibfnamefont {Gilbert}\ \bibnamefont
  {Strang}},\ }\href@noop {} {\emph {\bibinfo {title} {Linear Algebra and Its
  Applications}}},\ \bibinfo {edition} {4th}\ ed.\ (\bibinfo  {publisher}
  {{Cengage Learning}},\ \bibinfo {address} {{Belmont, CA}},\ \bibinfo {year}
  {2006})\BibitemShut {NoStop}%
\bibitem [{\citenamefont {Wolf}(2004)}]{wolf2004}%
  \BibitemOpen
  \bibfield  {author} {\bibinfo {author} {\bibfnamefont {Kurt~Bernardo}\
  \bibnamefont {Wolf}},\ }\href@noop {} {\emph {\bibinfo {title} {Geometric
  Optics on Phase Space}}}\ (\bibinfo  {publisher} {Springer-Verlag},\ \bibinfo
  {address} {Berlin},\ \bibinfo {year} {2004})\BibitemShut {NoStop}%
\bibitem [{\citenamefont {Pl{\"u}ker}(1865)}]{plueker1865}%
  \BibitemOpen
  \bibfield  {author} {\bibinfo {author} {\bibfnamefont {Julius}\ \bibnamefont
  {Pl{\"u}ker}},\ }\bibfield  {title} {\enquote {\bibinfo {title} {On a new
  geometry of space},}\ }\href {\doibase 10.1098/rspl.1865.0014} {\bibfield
  {journal} {\bibinfo  {journal} {Proceedings of the Royal Society of London}\
  }\textbf {\bibinfo {volume} {14}},\ \bibinfo {pages} {53--58} (\bibinfo
  {year} {1865})}\BibitemShut {NoStop}%
\bibitem [{\citenamefont {Hestenes}(2003)}]{hestenes2003}%
  \BibitemOpen
  \bibfield  {author} {\bibinfo {author} {\bibfnamefont {David}\ \bibnamefont
  {Hestenes}},\ }\bibfield  {title} {\enquote {\bibinfo {title} {{Oersted Medal
  Lecture} 2002: {Reforming} the mathematical language of physics},}\ }\href
  {\doibase 10.1119/1.1522700} {\bibfield  {journal} {\bibinfo  {journal}
  {American Journal of Physics}\ }\textbf {\bibinfo {volume} {71}},\ \bibinfo
  {pages} {104--121} (\bibinfo {year} {2003})}\BibitemShut {NoStop}%
\bibitem [{\citenamefont {Macdonald}(2011)}]{macdonald2011}%
  \BibitemOpen
  \bibfield  {author} {\bibinfo {author} {\bibfnamefont {Alan}\ \bibnamefont
  {Macdonald}},\ }\href@noop {} {\emph {\bibinfo {title} {Linear and Geometric
  Algebra}}}\ (\bibinfo  {publisher} {CreateSpace Independent Publishing
  Platform},\ \bibinfo {year} {2011})\BibitemShut {NoStop}%
\bibitem [{\citenamefont {Doran}\ \emph {et~al.}(1993)\citenamefont {Doran},
  \citenamefont {Hestenes}, \citenamefont {Sommen},\ and\ \citenamefont
  {Van~Acker}}]{doran1993}%
  \BibitemOpen
  \bibfield  {author} {\bibinfo {author} {\bibfnamefont {C.}~\bibnamefont
  {Doran}}, \bibinfo {author} {\bibfnamefont {D.}~\bibnamefont {Hestenes}},
  \bibinfo {author} {\bibfnamefont {F.}~\bibnamefont {Sommen}}, \ and\ \bibinfo
  {author} {\bibfnamefont {N.}~\bibnamefont {Van~Acker}},\ }\bibfield  {title}
  {\enquote {\bibinfo {title} {Lie groups as spin groups},}\ }\href {\doibase
  10.1063/1.530050} {\bibfield  {journal} {\bibinfo  {journal} {Journal of
  Mathematical Physics}\ }\textbf {\bibinfo {volume} {34}},\ \bibinfo {pages}
  {3642--3669} (\bibinfo {year} {1993})}\BibitemShut {NoStop}%
\bibitem [{\citenamefont {Dorst}(2016)}]{dorst2016}%
  \BibitemOpen
  \bibfield  {author} {\bibinfo {author} {\bibfnamefont {Leo}\ \bibnamefont
  {Dorst}},\ }\bibfield  {title} {\enquote {\bibinfo {title} {3d oriented
  projective geometry through versors of ${R}^{(3,3)}$},}\ }\href {\doibase
  10.1007/s00006-015-0625-y} {\bibfield  {journal} {\bibinfo  {journal}
  {Advances in Applied Clifford Algebras}\ }\textbf {\bibinfo {volume} {26}},\
  \bibinfo {pages} {1137--1172} (\bibinfo {year} {2016})}\BibitemShut {NoStop}%
\bibitem [{\citenamefont {Sugon}\ and\ \citenamefont
  {McNamara}(2003)}]{sugon2003}%
  \BibitemOpen
  \bibfield  {author} {\bibinfo {author} {\bibfnamefont {Quirino~M.}\
  \bibnamefont {Sugon}}\ and\ \bibinfo {author} {\bibfnamefont {Daniel~J.}\
  \bibnamefont {McNamara}},\ }\bibfield  {title} {\enquote {\bibinfo {title} {A
  geometric algebra reformulation of geometric optics},}\ }\href {\doibase
  10.1119/1.1621029} {\bibfield  {journal} {\bibinfo  {journal} {American
  Journal of Physics}\ }\textbf {\bibinfo {volume} {72}},\ \bibinfo {pages}
  {92--97} (\bibinfo {year} {2003})}\BibitemShut {NoStop}%
\bibitem [{\citenamefont {Sugon}\ and\ \citenamefont
  {McNamara}(2006)}]{sugon2006}%
  \BibitemOpen
  \bibfield  {author} {\bibinfo {author} {\bibfnamefont {Quirino~M.}\
  \bibnamefont {Sugon}}\ and\ \bibinfo {author} {\bibfnamefont {Daniel~J.}\
  \bibnamefont {McNamara}},\ }\bibfield  {title} {\enquote {\bibinfo {title}
  {Ray tracing in spherical interfaces using geometric algebra},}\ }\href
  {\doibase 10.1016/S1076-5670(05)39003-3} {\bibfield  {journal} {\bibinfo
  {journal} {Advances in Imaging and Electron Physics}\ }\textbf {\bibinfo
  {volume} {139}},\ \bibinfo {pages} {179--224} (\bibinfo {year}
  {2006})}\BibitemShut {NoStop}%
\bibitem [{\citenamefont {Sugon}\ and\ \citenamefont
  {McNamara}(2008)}]{sugon2008a}%
  \BibitemOpen
  \bibfield  {author} {\bibinfo {author} {\bibfnamefont {Quirino~M.}\
  \bibnamefont {Sugon}}\ and\ \bibinfo {author} {\bibfnamefont {Daniel~J.}\
  \bibnamefont {McNamara}},\ }\href@noop {} {\enquote {\bibinfo {title}
  {Paraxial meridional ray tracing equations from the unified
  reflection-refraction law via geometric algebra},}\ } (\bibinfo {year}
  {2008}),\ \Eprint {http://arxiv.org/abs/0810.5224} {arXiv:0810.5224}
  \BibitemShut {NoStop}%
\bibitem [{\citenamefont {Sugon~Jr.}\ and\ \citenamefont
  {McNamara}(2008)}]{sugon2008b}%
  \BibitemOpen
  \bibfield  {author} {\bibinfo {author} {\bibfnamefont {Quirino~M.}\
  \bibnamefont {Sugon~Jr.}}\ and\ \bibinfo {author} {\bibfnamefont {Daniel~J.}\
  \bibnamefont {McNamara}},\ }\href@noop {} {\enquote {\bibinfo {title}
  {Poisson commutator-anticommutator brackets for ray tracing and longitudinal
  imaging via geometric algebra},}\ } (\bibinfo {year} {2008}),\ \Eprint
  {http://arxiv.org/abs/0812.2979} {arXiv:0812.2979} \BibitemShut {NoStop}%
\bibitem [{\citenamefont {Pottman}\ and\ \citenamefont
  {Wallner}(2001)}]{pottman2001}%
  \BibitemOpen
  \bibfield  {author} {\bibinfo {author} {\bibfnamefont {H.}~\bibnamefont
  {Pottman}}\ and\ \bibinfo {author} {\bibfnamefont {J.}~\bibnamefont
  {Wallner}},\ }\href@noop {} {\emph {\bibinfo {title} {Computational Line
  Geometry}}}\ (\bibinfo  {publisher} {{Springer}},\ \bibinfo {address} {{New
  York}},\ \bibinfo {year} {2001})\BibitemShut {NoStop}%
\bibitem [{\citenamefont {Arnaud}(1985)}]{arnaud1985}%
  \BibitemOpen
  \bibfield  {author} {\bibinfo {author} {\bibfnamefont {Jacques}\ \bibnamefont
  {Arnaud}},\ }\bibfield  {title} {\enquote {\bibinfo {title} {Representation
  of {{Gaussian}} beams by complex rays},}\ }\href {\doibase
  10.1364/AO.24.000538} {\bibfield  {journal} {\bibinfo  {journal} {Applied
  Optics}\ }\textbf {\bibinfo {volume} {24}},\ \bibinfo {pages} {538} (\bibinfo
  {year} {1985})}\BibitemShut {NoStop}%
\bibitem [{\citenamefont {Colbourne}(2014)}]{colbourne2014}%
  \BibitemOpen
  \bibfield  {author} {\bibinfo {author} {\bibfnamefont {Paul~D.}\ \bibnamefont
  {Colbourne}},\ }\bibfield  {title} {\enquote {\bibinfo {title} {Generally
  astigmatic {{Gaussian}} beam representation and optimization using skew
  rays},}\ }in\ \href {\doibase 10.1117/12.2071105} {\emph {\bibinfo
  {booktitle} {SPIE Proceedings}}},\ Vol.\ \bibinfo {volume} {9293},\ \bibinfo
  {editor} {edited by\ \bibinfo {editor} {\bibfnamefont {Mariana}\ \bibnamefont
  {Figueiro}}, \bibinfo {editor} {\bibfnamefont {Scott}\ \bibnamefont
  {Lerner}}, \bibinfo {editor} {\bibfnamefont {Julius}\ \bibnamefont
  {Muschaweck}}, \ and\ \bibinfo {editor} {\bibfnamefont {John}\ \bibnamefont
  {Rogers}}}\ (\bibinfo {address} {{Kohala Coast, Hawaii, United States}},\
  \bibinfo {year} {2014})\ p.\ \bibinfo {pages} {92931S}\BibitemShut {NoStop}%
\bibitem [{\citenamefont {Collins}(1970)}]{collins1970}%
  \BibitemOpen
  \bibfield  {author} {\bibinfo {author} {\bibfnamefont {Stuart~A.}\
  \bibnamefont {Collins}},\ }\bibfield  {title} {\enquote {\bibinfo {title}
  {Lens-system diffraction integral written in terms of matrix optics},}\
  }\href {\doibase 10.1364/JOSA.60.001168} {\bibfield  {journal} {\bibinfo
  {journal} {Journal of the Optical Society of America}\ }\textbf {\bibinfo
  {volume} {60}},\ \bibinfo {pages} {1168--1177} (\bibinfo {year}
  {1970})}\BibitemShut {NoStop}%
\bibitem [{\citenamefont {Bandres}\ and\ \citenamefont
  {Guizar-Sicairos}(2009)}]{bandres2009}%
  \BibitemOpen
  \bibfield  {author} {\bibinfo {author} {\bibfnamefont {Miguel~A.}\
  \bibnamefont {Bandres}}\ and\ \bibinfo {author} {\bibfnamefont {Manuel}\
  \bibnamefont {Guizar-Sicairos}},\ }\bibfield  {title} {\enquote {\bibinfo
  {title} {Paraxial group},}\ }\href {\doibase 10.1364/OL.34.000013} {\bibfield
   {journal} {\bibinfo  {journal} {Optics Letters}\ }\textbf {\bibinfo {volume}
  {34}},\ \bibinfo {pages} {13} (\bibinfo {year} {2009})}\BibitemShut {NoStop}%
\end{thebibliography}%

\ifarXiv
    

\section*{Supplementary material}

# Supplemental Material for "Beyond the ABCDs: A better matrix method for geometric optics by using homogeneous coordinates"

Theodore A. Corcovilos*

*Department of Physics, Duquesne University, Pittsburgh, PA 15282*

(Dated: 14 February 2023)

## CONTENTS

## I. ALGEBRA RULES FOR RAYS AND POINTS IN HOMOGENEOUS COORDINATES

In Section 6 of the main text, we define our homogeneous representation for rays $r$ and points $p$. Here we present additional relationships between rays and points that provide computational shortcuts to some common tasks. We sketch out simplified derivations for these results, glossing over details such as the special cases of infinite points and rays. For rigorous proofs, see Ref.[1] and for examples see Ref.[2].

As a reminder, rays are expressed as column vectors $r = (c, a, b)^T$ corresponding the the line satisfying the equation $ax + by + c = 0$, and points are expressed as column vectors $p = [w, x, y]^T$, corresponding the the physical point $[x/w, y/w]$. Furthermore, we have the following normalization conventions: points are normalized if $w = 1$ and rays are normalized if $a^2 + b^2 = 1$. With these definitions, we can state the following Rules:

1. A ray passes through a point if and only if their vector dot product is zero: $r \cdot p = 0$.

2. The intersection point of two rays is the vector cross product of their ray vectors: $p = r_1 \times r_2$.

3. The ray vector connecting two points is the vector cross product of the point vectors: $r = p_1 \times p_2$.

4. Given three normalized finite points, the oriented area of their enclosed triangle is $A = \frac{1}{2}(p_1 \times p_2) \cdot p_3$. (Positive if the points are listed going counterclockwise around the triangle.) If the three points are colinear, then $(p_1 \times p_2) \cdot p_3 = 0$.

5. If the vectors $r$ and $p$ are normalized according to the conventions above then $r \cdot p$ is the signed distance from the point to the ray: positive if the point is on the left side of the ray and negative if the point is on the right side of the ray.

The above expressions work for two-dimensional rays and points in our homogeneous representation.

Rule 1 can be seen by construction:

$$0 = r\dot{p} = cw + ax + by,$$
$$= c + a\left(\frac{x}{w}\right) + b\left(\frac{y}{w}\right)$$

is an equation for a line passing through the point $[x, y]$.

For Rule 2, the requiring both rays $r_1, r_2$ to pass through the point $p$ gives us a pair of linear equations:

$$0 = r_1 \cdot p = c_1 w + a_1 x + b_1 y,$$
$$0 = r_2 \cdot p = c_2 w + a_2 x + b_2 y.$$

Rearranging...

$$a_1 x + b_1 y = -c_1 w,$$
$$a_2 x + b_2 y = -c_2 w.$$

This system of equations can be solved using Kramer's Rule:

$$x = \frac{\begin{vmatrix} -c_1 w & b_1 \\ -c_2 w & b_2 \end{vmatrix}}{\begin{vmatrix} a_1 & b_1 \\ a_2 & b_2 \end{vmatrix}} = w\frac{b_1 c_2 - b_2 c_1}{a_1 b_2 - a_2 b_1}, \quad y = \frac{\begin{vmatrix} a_1 & -c_1 w \\ a_2 & -c_2 w \end{vmatrix}}{\begin{vmatrix} a_1 & b_1 \\ a_2 & b_2 \end{vmatrix}} = w\frac{a_2 c_1 - a_1 c_2}{a_1 b_2 - a_2 b_1}. \tag{1}$$

Compare this to the claim that the vector cross product of the two ray vectors equals their intersection point:

$$r_1 \times r_2 = [a_1 b_2 - a_2 b_1, \quad b_1 c_2 - b_2 c_1, \quad a_2 c_1 - a_1 c_2]^T,$$
$$= [w, x, y]^T.$$

Substituting these values of $w, x, y$ into Eq. (1) above verifies the proposition.

The proof of Rule 3 is similar to Rule 2. The ray $r = (c, a, b)^T$ must contain both points $p_1, p_2$:

$$0 = r \cdot p_1 = cw_1 + ax_1 + by_1,$$
$$0 = r \cdot p_2 = cw_2 + ax_2 + by_2.$$

Rearranging...

$$x_1 a + y_1 b = -cw_1,$$
$$x_2 a + y_2 b = -cw_2.$$

As above, we solve for $a, b$ and compare the expression with that for $p_1 \times p_2$ to verify the claim.

For Rule 4, first consider the colinear case. Define $r_{12} = p_1 \times p_2$, which from Rule 3 is the ray connecting

points $p_1$ and $p_2$. If point $p_3$ also lies on $r_{12}$, then $r_{12} \cdot p_3 = 0$ from Rule 1. Putting this together, we get that three colinear points obey $(p_1 \times p_2) \cdot p_3 = 0$.

For the area rule, without loss of generality pick $p_1 = [1, 0, 0]^T$ to be the coordinate origin and $p_2 = [1, x_2, 0]^T$ to be on the $x$-axis. (This can be done by translating and rotating the coordinate axes.) The third point $p_3 = [1, x_3, y_3]^T$ is free. (All points are normalized using the convention $w = 1$.) The base of this triangle can be taken as side along the $x$-axis and the height as the $y$ coordinate of $p_3$, giving area $A = x_2 y_3 / 2$ (keeping the signs of the coordinates).

The formula for Rule 4 gives

$$A = \frac{1}{2}(p_1 \times p_2) \cdot p_3 = \frac{1}{2} \begin{vmatrix} 1 & 1 & 1 \\ 0 & x_2 & x_3 \\ 0 & 0 & y_3 \end{vmatrix} = \frac{1}{2} x_2 y_3,$$

where the determinant is a shortcut for calculating the triple product. This agrees with the expression for the area given above.

Rule 5 follows the pattern of Rule 4, where one side of the triangle is represented by a normalized ray, rather than a pair of points. The area of this triangle is $A = \frac{1}{2} r \cdot p = \frac{1}{2} bh$. The base $b$ of this triangle has unit length and the height $h$ of the resulting triangle is the perpendicular distance from the line to the point. Therefore, $h = r \cdot p$.

The above Rules work for two-dimensional rays and points in our homogeneous representation. For generalizations to three dimensions (and more) see Refs.[1–6]

## II. DERIVATION OF THE POINT TRANSFER MATRIX USING GRASSMANN EXTERIOR ALGEBRA

This section will introduce the machinery needed to derive the point transfer matrices more directly using Grassmann exterior algebra[4,7], by applying it to map the ray transfer transformation $M$ in the vector space of rays to a corresponding transformation $\overline{M}$ in the vector space of (homogeneous) points. The inclusion of exterior algebra invites an expanded set of interpretations. As an example of this, we'll close with a nearly trivial derivation of the Scheimpflug principle used, for example, to describe tilt-shift photography.

We define a graded vector algebra $\bigwedge^3 V$ with basis vectors $e_1, e_2, e_0$. The vector $e_2$ represents the nominal optical axis, $e_1$ represents the input plane of the optical system, and $e_0$ stands in for the ideal projective line at infinity (Fig. 1). Higher grades of elements are built up using Grassmann exterior (wedge) products of vectors, which obey simply

$$a \wedge b = -b \wedge a.$$

This implies $a \wedge a = 0$. More generally, the exterior product of any set of linearly dependent vectors will be zero.

Other than this simplification for linearly dependent vectors, in general the exterior products do not reduce further.

The unit bivectors $e_{01}, e_{20}, e_{12}$, using the abbreviation $e_{ij} = e_i \wedge e_j$, represent the $+y$ direction (or ideal point along the $+y$ axis), $+x$ direction, and origin, respectively. The trivector $e_{012} = e_0 \wedge e_1 \wedge e_2$ represents the whole $xy$ plane. (The order of indices above is chosen to ensure consistent signs in what follows.)

More concretely, a line obeying the equation $ax + by + c = 0$ is described in this notation by the vector

$$r = ce_0 + ae_1 + be_2, \tag{2}$$

and a finite point with homogeneous coordinates $[w, x, y]^T$ is described by the bivector

$$P = we_{12} + xe_{20} + ye_{01}. \tag{3}$$

As with homogeneous coordinates, bivectors with $w = 0$ represent points infinitely far away in the corresponding direction.

The definitions of points and lines are congruent, including orientation, up to an overall scalar factor: e.g. $P$ and $cP$, with $c$ a positive scalar constant, represent the same geometric point and similarly with lines. For a general finite point, Eq. (3), we define the norm $\|P\|$ to be the coefficient of $e_{12}$. For a general line, Eq. (2), we will choose the norm to be $\|r\| = \sqrt{a^2 + b^2}$. We define normalized elements $\hat{a} \doteq a/\|a\|$. Notably, two lines $r_1$ and $r_2$ intersect at the point $r_1 \wedge r_2$ (the projective "meet" operation), which may be an infinite point in the case of parallel lines. If three lines $r_{1,2,3}$ meet at a common point then $r_1 \wedge r_2 \wedge r_3 = 0$. If we extend our Grassmann algebra to the Cayley-Grassmann algebra by inclusion of the regressive product $\vee$, then we get the full projective duality between lines and points in two dimensions.

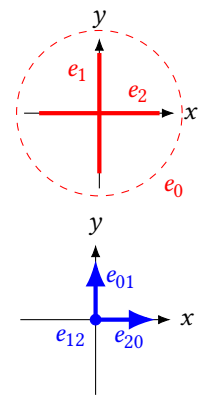

FIG. 1. (Color online only.) (Top) The unit vectors for our 2d projective space. $e_1$ and $e_2$ are lines perpendicular to the $x$ and $y$ axes, resp. $e_0$ is the ideal line infinitely far away from the origin. (Bottom) The unit bi-vectors for our space. $e_{20}$ and $e_{01}$ are the directions (ideal points) along the $x$ and $y$ axes, resp. $e_{12}$ represents the point at the origin.

The regressive product of two distinct points yields the line connecting them (the projective "join")[7]. If those two points are normalized, the norm of the resulting line gives the distance between the original points.

Our next task is to rewrite the ray transfer matrix equation,

$$\begin{pmatrix} h' \\ m' \end{pmatrix} = \begin{pmatrix} A & B \\ C & D \end{pmatrix} \begin{pmatrix} h \\ m \end{pmatrix}, \qquad (4)$$

which is also Eq. (1) of the main text, in this new notation. A ray with height $h$ at the $y$ axis and slope $m$ obeys the algebraic equation $y = mx + h$. Applying Eq. 2 we can associate this ray with a (un-normalized) line

$$r = -he_0 - me_1 + 1e_2,$$

which may be written as a column vector $r \doteq (-h, -m, 1)^T$ on the basis $\{e_0, e_1, e_2\}$ (note the order of the basis elements). Similarly, the outgoing ray is the column vector $r' \doteq (-h', -m', 1)^T$. A simple comparison shows that the ray transfer matrix equation Eq. (4) in this new representation becomes

$$r' = Mr,$$

$$\begin{pmatrix} -h' \\ -m' \\ 1 \end{pmatrix} = \begin{pmatrix} A & B & 0 \\ C & D & 0 \\ 0 & 0 & 1 \end{pmatrix} \begin{pmatrix} -h \\ -m \\ 1 \end{pmatrix}.$$

The purpose of this rewriting is to facilitate our next step: finding the matrix representation of the ray transfer matrix in the bivector basis. First, we'll take a small interlude to demonstrate an example.

The usual description of the ray transfer matrices is that they are valid in the limit of paraxial rays, defined as those with small slope and small ray heights, such that the ray tracing equations may be approximated as linear equations. The ray transfer matrices also hold in a complementary approximation: the approximation that the optical elements themselves are linear. For example, we could consider an ideal thin lens that exhibits no aberrations even for rays far from the axis.

With this proviso we can derive the Scheimpflug principle (Fig. 2). We'll consider the simplest case of an ideal thin lens. The Scheimpflug principle states when an extended object lies in a plane that is not normal to the optical axis, then the object plane, image plane, and the plane of the lens all meet at a point. Let the plane of the lens be the $y$-axis of our system, $e_1$. We place the object plane $p$ such that it crosses the optical axis at a distance $d$ from the lens and is tilted by an angle $\theta$ relative to normal. Because we are working in two dimensions, this is equivalent to the line

$$p = -d\cos\theta\, e_0 - \cos\theta\, e_1 + \sin\theta\, e_2.$$

We find the image plane, $p'$ by applying the thin lens ray transfer matrix to the object plane.

$$p' = \begin{pmatrix} 1 & 0 & 0 \\ -1/f & 1 & 0 \\ 0 & 0 & 1 \end{pmatrix} \begin{pmatrix} -d\cos\theta \\ -\cos\theta \\ \sin\theta \end{pmatrix} = \begin{pmatrix} -d\cos\theta \\ (d/f - 1)\cos\theta \\ \sin\theta \end{pmatrix}.$$

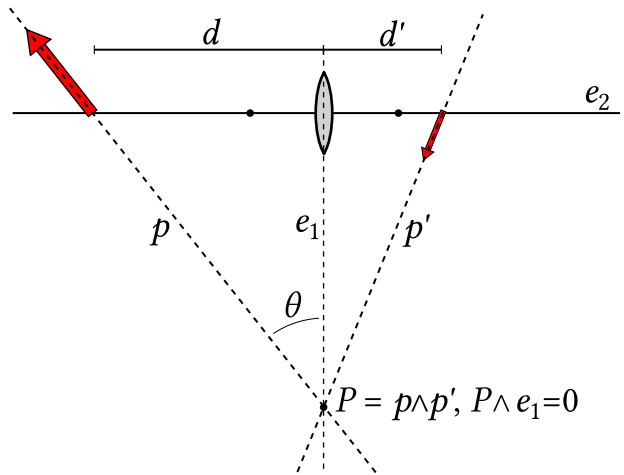

FIG. 2. (Color online only.) The Scheimpflug Principle. The plane $p$ of the tilted object (large red arrow) and the plane $p'$ of the resulting image (small red arrow) meet at a point $P$ in the plane of the lens ($e_1$).

With this, we can calculate the intersection point $P$ of $p$ and $p'$ as their wedge product:

$$P = p \wedge p' = -\frac{d^2\cos^2\theta}{f} e_{01} - \frac{d\sin\theta\cos\theta}{f} e_{12} \equiv \frac{d}{\tan\theta} e_{01} + e_{12},$$

where in the last step a common factor is removed to normalize the point, showing that the intersection point is indeed on the $y$ axis at a height of $d/\tan\theta$. We could also have checked that the three planes intersect by showing $p \wedge p' \wedge e_1 = 0$. We can also verify that the image plane crosses the optical axis ($e_2$) at the appropriate image point given by, e.g., Gauss's lens equation.

$$p' \wedge e_2 = d\cos\theta\, e_{20} + \left(\frac{d}{f} - 1\right) e_{12} \equiv \frac{df}{d-f} e_{20} + e_{12},$$

where the last step is normalization to show the expected location of the image point.

Because points are represented by bivectors in our algebra, we wish to find a representation for how bivectors transform in 2D optical systems. This will be given by the outermorphism of the linear operator. Taking a linear operator $\mathsf{L}$ acting on vectors (lines) in our space, we will extend it to higher grades by defining the outermorphism

$$\mathsf{L}(a \wedge b) = \mathsf{L}(a) \wedge \mathsf{L}(b). \qquad (5)$$

A matrix representation of $\mathsf{L}$ on the vector space of column vectors may be constructed by acting $\mathsf{L}$ on each of the basis elements $e_{0,1,2}$ and storing the results as the corresponding columns of a matrix. Similarly, a matrix representation on the bivector space can be constructed by acting $\mathsf{L}$ on each of the bivector basis elements through the outermorphism described above.

To begin, let's see how the ray transfer matrix $M$ transforms the basis vectors. The $e_1$ element transforms into

$$Me_1 \doteq \begin{pmatrix} A & B & 0 \\ C & D & 0 \\ 0 & 0 & 1 \end{pmatrix} \begin{pmatrix} 0 \\ 1 \\ 0 \end{pmatrix} = \begin{pmatrix} B \\ D \\ 0 \end{pmatrix},$$
$$= Be_0 + De_1.$$

Similarly, $Me_0 = Ae_0 + Ce_1$ and $Me_2 = e_2$. Not surprisingly, these are just the columns of $M$.

By the outermorphism property Eq. (5), we can find the matrix representation in the bivector space by operating on each basis bivector:

$$Me_{12} = Me_1 \wedge Me_2,$$
$$= (Be_0 + De_1) \wedge e_2 = De_{12} - Be_{20},$$
$$Me_{20} = -Ce_{12} + Ae_{20},$$
$$Me_{01} = (AD - BC)e_{01}.$$

Our (un-normalized) object-space point $P = we_{12} + xe_{20} + ye_{01}$ may be written as a homogeneous vector $[w, x, y]^T$ on the bivector basis $\{e_{12}, e_{20}, e_{01}\}$ (note order). The equations above for the transformation of the unit bivectors can be arranged as a matrix equation. The optical system generates a corresponding image-space point

$$P' = \begin{bmatrix} w' \\ x' \\ y' \end{bmatrix} = \begin{bmatrix} D & -C & 0 \\ -B & A & 0 \\ 0 & 0 & AD - BC \end{bmatrix} \begin{bmatrix} w \\ x \\ y \end{bmatrix}. \quad (6)$$

This confirms the result we presented earlier in Eq. (7) of the main text, adding the overall scalar factor.

---

[*] corcovilost@duq.edu

[1] Jorge Stolfi, *Oriented Projective Geometry: A Framework for Geometric Computations* (Academic Press, 2014).

[2] Matt Pharr, Wenzel Jakob, and Greg Humphreys, *Physically Based Rendering: From Theory to Implementation*, 3rd ed. (Morgan Kaufmann, Cambridge, MA, 2016).

[3] Chris Doran and A. N. Lasenby, *Geometric Algebra for Physicists* (Cambridge University Press, Cambridge, 2007).

[4] Sergei Winitzki, *Linear Algebra via Exterior Products*, v. 1.3 ed. (Lulu, 2020).

[5] Leo Dorst, Daniel Fontijne, and Stephen Mann, *Geometric Algebra for Computer Science* (Morgan Kaufmann, Amsterdam, 2007).

[6] Leo Dorst, "A guided tour to the plane-based geometric algebra PGA," (2020), intended as replacement for Ch. 11 of [5].

[7] John Browne, *Foundations*, Vol. 1 (CreateSpace, 2012) googlebooks:IuYIAwAAQBAJ.


\fi

  \end{document}